\documentclass{aa}
\usepackage{hyperref}

\usepackage{graphicx}
\usepackage{txfonts}
\usepackage{natbib}
\newcommand{\oii}{[\ion{O}{ii}]}
\newcommand{\oiiab}{[\ion{O}{ii}]$\lambda\lambda$3727,3729}
\newcommand{\oiia}{[\ion{O}{ii}]$\lambda$3727}
\newcommand{\oiib}{[\ion{O}{ii}]$\lambda$3729}
\newcommand{\oiii}{[\ion{O}{iii}]}
\newcommand{\oiiiab}{[\ion{O}{iii}]$\lambda\lambda$4959,5007}
\newcommand{\oiiia}{[\ion{O}{iii}]$\lambda$5007}
\newcommand{\oiiib}{[\ion{O}{iii}]$\lambda$4959}
\newcommand{\oiiic}{[\ion{O}{iii}]$\lambda$4363}

\newcommand{\mgii}{\ion{Mg}{ii}}
\newcommand{\neiii}{[\ion{Ne}{iii}]$\lambda 3869$}
\newcommand{\neiiia}{[\ion{Ne}{iii}]}
\newcommand{\nev}{[\ion{Ne}{v}]$\lambda 3426$}
\newcommand{\neva}{[\ion{Ne}{v}]}
\newcommand{\hb}{H$\beta$}
\newcommand{\hg}{H$\gamma$}
\newcommand{\hd}{H$\delta$}

\newcommand{\ha}{H$\alpha$}
\usepackage{color}
\usepackage{amstext}

\begin{document} 

   \title{Ionised gas structure of 100 kpc in an over-dense region of the galaxy group COSMOS-Gr30 at z $\sim$ 0.7\thanks{Based on observations made with ESO telescopes at the Paranal Observatory under programs 094.A-0247 and 095.A-0118.}}
   \titlerunning{Ionised gas structure of 100 kpc in an over-dense region at z $\sim$ 0.7} 

  \author{
    B. Epinat\inst{1, 2}
          \fnmsep\thanks{\email{benoit.epinat@lam.fr}}
  \and
    T. Contini\inst{1}
  \and
    H. Finley \inst{1}
  \and
    L. A. Boogaard\inst{3}
  \and
    A. Guérou\inst{1, 4}
  \and
    J. Brinchmann\inst{3, 5}
  \and
    D. Carton\inst{6}
  \and
    L. Michel-Dansac\inst{6}
  \and
    R. Bacon\inst{6}
  \and
    S. Cantalupo\inst{7}
  \and
    M. Carollo\inst{7}
  \and
    S. Hamer\inst{6}
  \and
    W. Kollatschny\inst{8}
  \and
    D. Krajnovi\'{c}\inst{9}
  \and
    R. A. Marino\inst{7}
  \and
    J. Richard\inst{6}
  \and
    G. Soucail\inst{1}
  \and
    P. M. Weilbacher\inst{9}
  \and
    L. Wisotzki\inst{9}
  }

   \institute{Institut de Recherche en Astrophysique et Planétologie (IRAP), Université de Toulouse, CNRS, UPS, CNES, Toulouse, France
    \and
      Aix Marseille Univ, CNRS, LAM, Laboratoire d'Astrophysique de Marseille, Marseille, France
    \and
      Leiden Observatory, Leiden University, P.O. Box 9513, 2300 RA Leiden, The Netherlands
    \and
      European Southern Observatory, Karl-Schwarzschild-Str. 2, D-85748 Garching, Germany
    \and
      Instituto de Astrof{\'\i}sica e Ci{\^e}ncias do Espaço, Universidade do Porto, CAUP, Rua das Estrelas, PT4150-762 Porto, Portugal
    \and    
      Univ Lyon, Univ Lyon1, Ens de Lyon, CNRS, Centre de Recherche Astrophysique de Lyon UMR5574, F-69230, Saint-Genis-Laval, France
    \and
      Institute for Astronomy, Department of Physics, ETH Zürich, Wolfgang-Pauli-Strasse, 27, CH-8093, Zürich, Switzerland
    \and
      Institut fur Astrophysik, Universitat Gottingen, Friedrich-Hund Platz 1, 37077, Göttingen, Germany
    \and
      Leibniz-Institut für Astrophysik Potsdam (AIP), An der Sternwarte 16, D-14482 Potsdam, Germany
      }

   \date{Received September 1, 2017; accepted October 27, 2017}

  \abstract
   {We report the discovery of a $10^4$~kpc$^2$ gaseous structure detected in \oiiab\ in an over-dense region of the 
   COSMOS-Gr30 galaxy group
   at $z\sim 0.725$ with deep MUSE Guaranteed Time Observations.
   We estimate the total amount of diffuse ionised gas to be of the order of $(\sim 5 \pm 3) \times10^{10}$~M$_\odot$ and explore its physical properties to understand its origin and the source(s) of the ionisation.
   The MUSE data allow the identification of a dozen  group members that are embedded in this structure through emission and absorption lines.
   We extracted spectra from small apertures defined for both the diffuse ionised gas and the galaxies. We investigated the kinematics and ionisation properties of the various galaxies and extended gas regions through line diagnostics (R23, O32, and \oiii/\hb) that are available within the MUSE wavelength range.
   We compared these diagnostics to photo-ionisation models and shock models.
   The structure is divided into two kinematically distinct sub-structures. The most extended sub-structure of ionised gas is likely rotating around a massive galaxy and displays filamentary patterns that
link some galaxies.
   The second sub-structure links another massive galaxy that
hosts an active galactic nucleus (AGN) to a low-mass galaxy, but it also extends orthogonally to the AGN host disc over $\sim 35$~kpc. This extent is likely ionised by the AGN itself. The location of small diffuse regions in the R23 vs. O32 diagram is compatible with photo-ionisation. However, the location of three of these regions in this diagram (low O32, high R23) can also be explained by shocks, which is supported by their high velocity dispersions. One edge-on galaxy shares the same properties and may be a source of shocks.
   Regardless of the hypothesis, the extended gas seems to be non-primordial.
   We favour a scenario where the gas has been extracted from galaxies by tidal forces and AGN triggered by interactions between at least the two sub-structures.}


   \maketitle
%

\section{Introduction}
\label{introduction}

The environment is expected to play a major role in the galaxy mass assembly processes, star formation quenching, and morphological transformation of galaxies. 
At any cosmological time, dense environments indeed include a larger fraction of passive galaxies \citep{Cucciati:2010}.
Furthermore, a range of studies at low \citep[e.g.][]{Peng:2010, Cibinel:2013} and intermediate redshift \citep[e.g.][]{Cucciati:2006, McGee:2011, Muzzin:2013} have shown that the quenching of star formation mainly depends on the galaxy mass, and, particularly in the case of low-mass galaxies (M$_* \lesssim 10^{9.5}$~M$_\odot$), on their environment and their location in the dark matter halo.
Environmental quenching is thought to be prominent at $z<0.5$ \citep{Peng:2010}, but the turnover is not yet well constrained. It also seems that the environmental processes that turn off star formation operate on a fairly long timescale (a few Gyr, \citealp{Cibinel:2013, Wetzel:2013}), implying that the galaxies in $z\sim 0.1$ groups that have recently turned off their star formation will have started this process at $z>0.5$.
These processes may be driven by violent mechanisms that are
linked to the more frequent interactions between galaxies. These interactions can extract gas from galaxies, affect reservoirs of gas around galaxies, and may produce extended gas structures.


In the nearby Universe, the cool cores seen in the X-ray emission from galaxy groups and clusters can result in the rapid deposition of gas in the central regions of these dense regions \citep[see][for reviews]{Fabian:1994, McNamara:2007}. This process commonly leads to the production of ionised gas structures that surround the brightest cluster galaxies (BCGs) that are found at the centres of these clusters \citep{Heckman:1989}, which are usually filamentary and can extend over several tens of kpc \citep[e.g.][]{Conselice:2001, Hamer:2016}. These ionised gas regions trace the structure of a multiphase gas reservoir, which is also seen in X-ray \citep{Fabian:2008} and the cold \citep{Salome:2011} and warm \citep{Wilman:2009} molecular gas emission. This requires an unusual heating source, possibly by fast-ionising particles \citep{Ferland:2009}, to explain the observed spectra.

Local compact groups also show evidence of strong interactions, similar to the interactions that are believed to play an important role in driving galaxy evolution at higher redshift. Stephan's Quintet is one of the most spectacular of such groups. In this group, a 35 kpc intergalactic filamentary structure is observed in radio continuum as well as in optical emission lines and X-rays. The extended gas in this structure appears to be ionised by shocks \citep{Appleton:2006, Appleton:2013, Konstantopoulos:2014, Guillard:2012} that are triggered by the gravitational interactions. Integral field spectroscopy has been used to study the kinematics and ionisation source in both galaxies \citep{Rosales-Ortega:2010, Rodriguez-Baras:2014} and intergalactic shocked gas \citep{Iglesias-Paramo:2012} of Stephan's Quintet. 
The studies found that the ionisation of the gas in some galaxies is jointly due to active galactic nuclei (AGN), stellar photo-ionisation, and shocks. They also found that the intergalactic large-scale shock region contains a high-velocity shocked component with a low metallicity and another component with a low velocity and solar metallicity.

The impact of strong gravitational interactions can also be probed through local ultra-luminous infrared galaxies (ULIRGs). They are starburst galaxies that are often induced by mergers of gas-rich galaxies. 
For this reason, they may resemble $z>0.5$ interacting galaxies.
Some studies have shown that ULIRGs display shocks that account for a considerable fraction of flux from the ionised gas \citep{Rich:2015}. These shocks are likely driven by interactions. During a merger, gas flows inwards as a result of tidal forces, generating bursts of star formation and AGN activity. These events can produce galactic outflows and shocks in the interstellar medium and beyond.

During interactions, some gas can be ejected from galaxies. As part of the SDSS-IV/MaNGA survey \citep{Bundy:2015}, \citet{Lin:2017} reported the detection of large \ha\ blobs with no associated optical counterpart 8 kpc away from one component of a dry merger at $z\sim 0.03$. These blobs may be ionised by a combination of massive young stars and AGN and may result from an AGN outflow or simply be associated with a low surface brightness galaxy.

At higher redshift, where star formation is enhanced with respect to the local Universe and where groups and clusters are forming, extended gas structures may be more frequent. However, only a few intergalactic ionised gas nebulae have been observed so far.
At $0.1<z<0.4$, \citet{Tumlinson:2011} observed large (up to 150 kpc) oxygen-rich halos around star-forming galaxies through quasar absorption in the ultraviolet. These reservoirs of gas and metals seem to be removed or transformed during star formation quenching that may occur in dense environments. The circumgalactic medium in these structures may be ionised by collisions or radiative cooling \citep[e.g.][]{Werk:2016, McQuinn:2017}. It may also be structured gas clouds that are photo-ionised by local high-energy sources.

In galaxy clusters at $z\sim 0.5$, emission lines associated with intracluster light can also be observed. \citet{Adami:2016} detected a region with \oii\ and \oiii\ emission lines, but no visible continuum counterpart. The ionisation properties of this source are, however, compatible with a low surface brightness galaxy, and would therefore not necessarily be due to an interaction inside the cluster.

At $z\sim 1$, \ha, \oiiab\ and \oiiia\ blobs have been observed around galaxies \citep[e.g.][]{Yuma:2013, Harikane:2014, Yuma:2017}. Such blobs can be as large as 75 kpc, but seem to be mainly associated with AGN outflows.

At redshifts $z>2$, giant fluorescent extended Ly$\alpha$ nebulae have been reported around bright quasars \citep[e.g.][]{Cantalupo:2014, Borisova:2016, Hennawi:2015, North:2017}. These extended nebulae may be located at the intersection of cosmic web filaments. Recently, a giant Ly$\alpha$ nebula has been observed in a galaxy over-density at $z\sim 2.3$ \citep{Cai:2017}. It might be powered by shocks due to an AGN-driven outflow and/or photo-ionisation by a strongly obscured source.

The Spiderweb Galaxy \citep{Miley:2006}, located in a proto-cluster of galaxies at $z\sim 2.2$ \citep{Kuiper:2011}, has a large reservoir of molecular gas \citep{Emonts:2013}. Molecular gas is detected inside the galaxy, but also in satellite galaxies and in the intracluster medium. This gas may fuel star formation that is
seen in the ultraviolet \citep{Hatch:2008, Emonts:2016}. However, extended ionised gas has not yet been observed in this galaxy.

At redshift $z\sim 3$, the discovery by \citet{Steidel:2000} of two large Ly$\alpha$ blobs in a protocluster led to deeper searches for such giant blobs that are not necessarily associated with quasars \citep{Matsuda:2004, Matsuda:2011} and to integral field spectroscopy follow-ups \citep[e.g. ][]{Weijmans:2010} to map the distribution more accurately. \citet{Matsuda:2011} found that these Ly$\alpha$ blobs may preferentially be associated with high-density environments and be related to large-scale outflows powered by either intense starburst or AGN activities.
However, Ly$\alpha$ emission has also been detected with MUSE around low- to intermediate-mass galaxies at $z>3$ \citep{Wisotzki:2016, Leclercq:2017} and extends over several tens of kiloparsecs outside of the galaxies.

In previous studies of extended ionised gas structures that were
discovered at various redshifts, integral field spectroscopy has been commonly used to map spectral and kinematics properties of these structures using various lines without any assumption on their distribution \citep[e.g. ][]{Cheung:2016, Fensch:2016, Lin:2017, Weijmans:2010, Borisova:2016, Iglesias-Paramo:2012, Rodriguez-Baras:2014, Adami:2016}. Studying line diagnostics with integral field spectroscopy allows one to constrain the abundance and hence the origin of the gas, as well as the energetics of the ionising sources of these structures at any location.

In this paper, we report the serendipitous discovery of a new large ionised gas structure observed in an over-dense region of a galaxy group at redshift $z\sim 0.72$ with the Multi Unit Spectroscopic Explorer (MUSE) \citep{Bacon:2015}. This group is located in a larger scale structure that is identified as the COSMOS-Wall \citep{Iovino:2016}. This is a large filamentary structure that hosts a variety of environments, including a dense cluster, galaxy groups, filaments, less dense regions, and voids.
The ionised gas structure we report here extends between the galaxies. The sensitivity, field of view, spectral range, and resolution of the MUSE integral field spectrograph allows the mapping of both emission lines fluxes and kinematics. We use this capability here to infer the mass, origin, and sources of ionisation of the extended gas.

This paper is structured as follows. In section \ref{obs_datared} we introduce our dataset and the data reduction steps. 
In section \ref{extended} we show the ionised gas distribution and kinematics, identify the galaxies embedded in this structure, and extract some of their properties.
We present the results of a spectral analysis in section \ref{results}. Finally, we provide our interpretation of the observed structure in section \ref{interpretation} and summarise our results in section \ref{conclusion}.

Throughout the paper, we assume a $\Lambda$CDM cosmology with H$_0=70$~km~s$^{-1}$~Mpc$^{-1}$, $\Omega_M=0.3,$ and $\Omega_\Lambda=0.7$.

\section{Observations and data reduction}
\label{obs_datared}

The analysis presented in this paper relies on the exploitation of MUSE data cubes as well as multi-band images and photometry.

\subsection{MUSE observations and data reduction}
\label{observations}

The galaxy group COSMOS-Gr30 \citep{Knobel:2012} was observed during MUSE Guaranteed Time Observations (GTO) as part of a program focusing on the effect of environment on galaxy evolution processes over the past 8 Gyr (PI: T. Contini). A total of 10 hours of exposures time was obtained on this field, spread over three observing runs: 2 hours in December 2014 (Program ID 094.A-0247), 4 hours in April 2015 (095.A-0118), and 4 hours in May 2015 (095.A-0118). For each run, observing blocks (OBs) included four 900-second exposures with a field rotation of 90 degrees between each exposure, leading to a total of 40 exposures. Seeing variations between the OBs were measured from four identified stars in the field and range from 0.50'' to 1.05''.

We produced two main reductions of this field: one that includes only the best seeing (below 0.7'') observations for a total of 6.25 hours (25 exposures of 15 minutes), and another that amounts to 9.75 hours of data from 39 exposures with seeing below 0.9''. The latter excludes 1 exposure where the seeing was above 1'', since this is above the program requirements.
The 39-exposure reduction best reveals the extended ionised gas region that is the purpose of our analysis, therefore we focus on this data cube.

The reduction was performed using the MUSE standard pipeline version v1.1.90 \citep{Weilbacher:2012, Weilbacher:2014, Weilbacher:2015}.
To process the calibration files from each night of observation, we ran the bias, flat, wavelength calibration, and line-spread function (LSF) calculation steps for each integral field unit using standard parameters.
We then produced twilight cubes, since calibration files for this program include sets of eight sky flats from the night of observations or from a night within the same GTO run. 
Where possible, we selected sky flats observed when the temperature was similar ($\Delta T < 0.5$~$^\circ$C) to the science observations, and we calibrated each set of sky flats with an illumination exposure taken closest in time and temperature to the sky flats. The illumination exposure corrects for small temperature-dependent flux variations at the edge of slices that are due to flexure in the instrument.
We applied these steps to the standard star exposures.  

To process the science exposures, we likewise applied the calibration files from the corresponding night, including the LSF correction, twilight cube, and an appropriately chosen illumination exposure.
We used the same standard star, LTT3218, to calibrate the science exposures, except for the data from the December 2014 GTO run, where we used the standard star HD49798. 
We applied `model' sky subtraction to the science exposures with the default parameters, while including the LSF calibration files. 
Finally, we aligned the 39 science exposures (see the next paragraph for details about the astrometry), combined them to create the final data cube, and applied the Zurich Atmosphere Purge v0.6 software  \citep[ZAP;][]{Soto:2016} to further improve the sky subtraction.

To produce the final data cube, we used four stars in the field as anchors to match the MUSE astrometry and flux with the COSMOS catalogue values \citep{Capak:2007}.
For the astrometry, we first aligned the 39 science exposures to correct slight right ascension and declination offsets within the MUSE data.
We then measured the average right ascension and declination offsets between the MUSE and the reference coordinates from the four stars and applied this global astrometry offset when we combined the different exposures to create the final data cube.
For the flux calibration, we applied a constant scale factor to the final combined data cube such that the magnitudes of the stars in the MUSE field match the I-band magnitudes.
To determine the flux scale factor, we produced a 30-pixel sub-cube for each of the four stars and used the pipeline tool 'muse\_cube\_filter' to create Cousins I-band images for each star. 
From the I-band image, we fit a 2D Moffat profile to each star and summed the flux in the fitted Moffat profile to calculate the magnitude of the star in the MUSE cube. Taking the I-band magnitude as the reference, we determined the flux scale factor from the average magnitude offset.

The final data cube has a 0.2'' spatial sampling and a 1.25 \AA\ spectral sampling covering the spectral range from 4750 \AA\ to 9350 \AA, which are standard for MUSE nominal mode observations. Whereas the MUSE field of view is one square arcminute, the final combined data cube has a more extended shape, since we re-centred the field of view between the December 2014 and April/May 2015 observations to cover more galaxies within the group. 

After combining the 39 exposures, we measured the seeing in the final data cube from the four stars identified in the field. We modelled these stars with a 2D Gaussian function and computed the seeing as the average of their full-width at half-maximum (FWHM). We estimate the seeing to be 0.68'' at 7000 \AA. 

To characterise the line spread function (LSF), we used the results that \citet{Bacon:2017} and \citet{Guerou:2017} obtained. They analysed the LSF in two distinct MUSE fields, located in the Hubble Ultra Deep Field and the Hubble Deep Field South, and observed at multiple epochs using a set of 19 groups of one to ten sky lines spread over the MUSE wavelength range. They demonstrated that the MUSE LSF variation with wavelength is very stable. The two fields used in the two studies have the same acquisition pattern as the field studied here, which consists of four rotations of 90$^\circ$. The LSF FWHM is parameterised as
\begin{equation}
 {\rm FWHM} = \lambda^2 \times 5.866 \times 10^{-8} - \lambda \times 9.187 \times 10^{-4} + 6.040
 \label{lsf}
,\end{equation}
where FWHM and $\lambda$ are both in Angstroms.

We recall that the standard MUSE data reduction process also produces a variance cube.

\subsection{Ancillary dataset}
\label{ancillary}

Because this galaxy group is inside the COSMOS field \citep{Scoville:2007}, many ancillary data are available and already reduced. They are presented in the last COSMOS catalogue release \citep[COSMOS2015,][]{Laigle:2016}. These datasets include radio data from the VLA, infrared and far-infrared data from Spitzer (MIPS and IRAC) and Herschel (PACS and SPIRE), near-infrared and optical observations from the HST-ACS (F814W filter), the SDSS, the VIRCAM/VISTA camera (Y, J, H, K$_S$), the WIRCam/CFHT camera (H and K$_S$ bands) and the MegaCam/CFHT camera (U and I bands), the Kitt Peak National Observatory (K$_S$ band), the HSC/Subaru Y band and  SuprimeCam/Subaru camera (B, V, g, r, i, z filters as well as 14 medium- and narrow-band filters), ultraviolet data from the Galaxy Evolution Explorer (GALEX, near and far ultraviolet), and X-ray from Chandra and XMM. Part of these ancillary datasets are used for the analysis and interpretation of the extended gas region.

\section{Extended ionised gas structure in an over-dense region}
\label{extended}

\begin{figure*}
 \includegraphics[width=18cm]{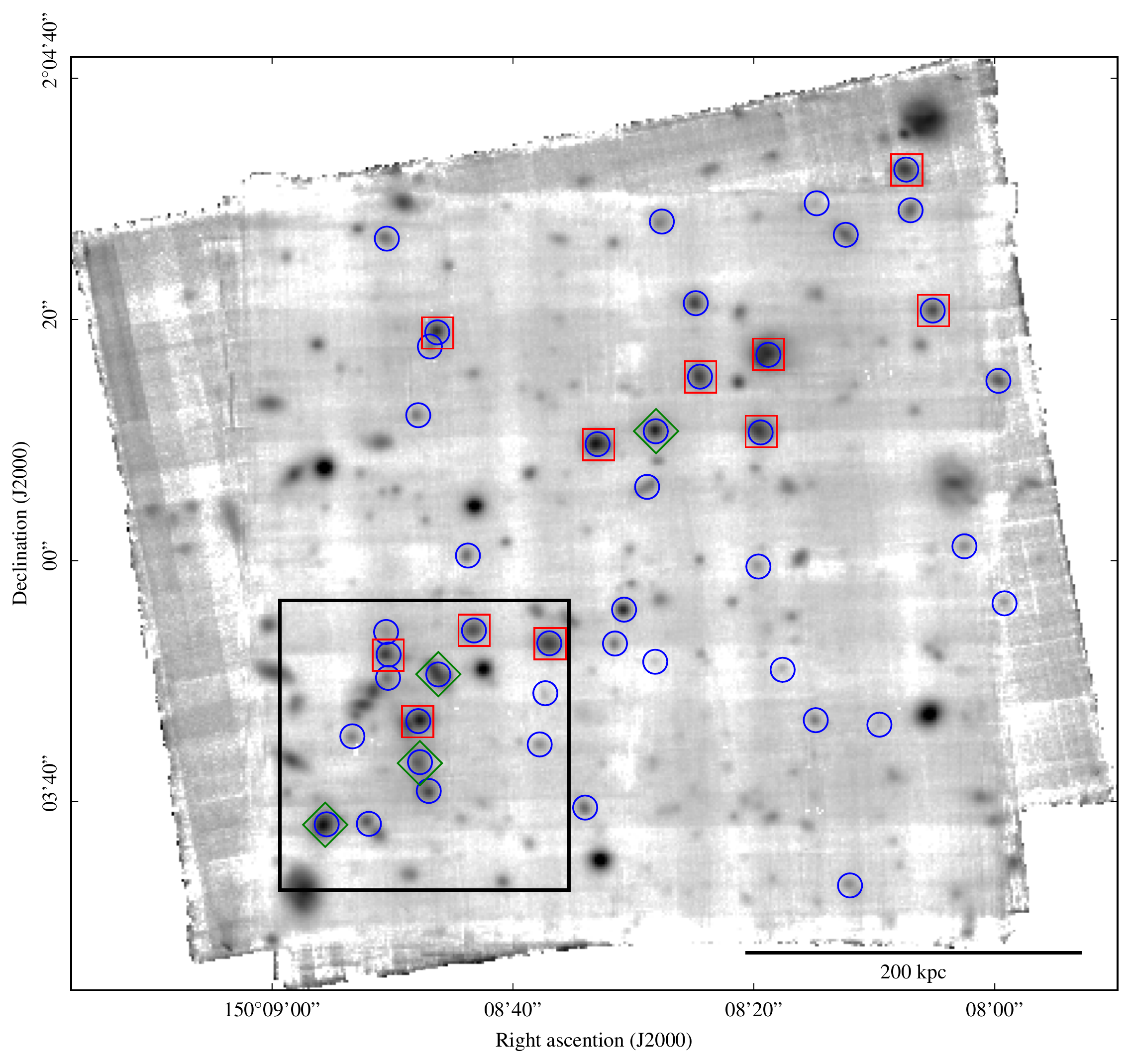}
 \caption{MUSE white-light image (cube averaged over all wavelengths, logarithmic scale, arbitrary unit) over the full field of view of the observations of COSMOS-Gr30. Spectroscopic and photometric group members from the zCOSMOS 20k group catalogue \citep{Knobel:2012} are indicated using red squares and green diamonds, respectively.
 The 44 group members that are unambiguously identified in the MUSE data cube are marked with blue circles.
 The black square represents the area studied in more detail in this paper. The physical scale at the redshift of the structure is indicated at the bottom right.}
\label{wli_largescale}
\end{figure*}

The galaxy group COSMOS-Gr30, which is at redshift $z\sim 0.725$, was targeted because of its particularly high number density. We have selected $z<1$ groups in the zCOSMOS 20k group catalogue \citep{Knobel:2012} with a large number of spectroscopically confirmed group members ($\geqslant 5$) inside one MUSE field of view. From the zCOSMOS 20k group catalogue, 11 spectroscopically confirmed members and 3 photometric candidates were expected in COSMOS-Gr30, making this group the richest of the catalogue. This group is in fact embedded in a larger structure identified as the COSMOS-Wall \citep{Iovino:2016}.
Using MUSE data, 44 group members were unambiguously spectroscopically identified with redshifts between $0.719 \leq z \leq 0.732$, increasing by a factor of four the number density of the group members in the field of view covered by our observations (cf. Figure \ref{wli_largescale}). These members include low-mass star-forming galaxies as well as passive galaxies without any emission line.
Of the 44 members, 25 have stellar masses higher than 10$^{9.5}$~M$_{\odot}$. The velocity dispersion of the group using these galaxies is 314~km~s$^{-1}$ and 402~km~s$^{-1}$ with not limit on the stellar mass.

In the following, we focus on the sub-area of the MUSE data cube highlighted in Figure \ref{wli_largescale}, which is where the ionised gas structure was found.

\subsection{Ionised gas structure and its distribution}
\label{ionised}

The ionised gas structure was serendipitously discovered during the extraction of ionised gas kinematics of galaxies from the MUSE combined data cube. We used the python code \emph{CAMEL}\footnote{\url{https://bitbucket.org/bepinat/camel.git}} described in \citet{Epinat:2012} to extract ionised gas kinematics of the structure by fitting emission lines in a sub-cube of $25 \times 25$ arcsec$^2$ field of view centred on $\text{RA}=150^\circ 08' 48''$ and $\text{Dec}=2^\circ 03' 44''$, fully covered by the 39 individual exposures (cf. section \ref{observations}). The extent of the field of view is equivalent to $\sim 180 \times 180$~kpc$^2$ at $z\sim0.725$.

Since the \oiiab\ doublet is the brightest emission line of the structure and since it is not affected by any strong absorption line, it was used alone to derive the structure kinematics, including line flux map, velocity field, and dispersion map.
Before extracting these maps, a spatial smoothing using a 2D Gaussian with a FWHM of two pixels was applied to the data cube in order to increase the signal-to-noise ratio.
For each spaxel, the \oii\ doublet was modelled by two Gaussian profiles sharing the same kinematics (same velocity and same velocity dispersion), but having distinct rest-frame wavelengths (3726.04 \AA\ and 3728.80 \AA) plus a constant continuum.
The variance data cube was used to weight each spectral element during line fitting in order to minimise the effect of noise,
which is mainly induced by sky lines.
In principle, the MUSE spectral resolution at 6430 \AA\ (LSF FWHM $\sim 2.55 \text{ \AA}$), corresponding to the observed wavelength of the \oii\ doublet for objects at $z\sim 0.725$, makes it possible to resolve the doublet at this redshift ($\Delta \lambda \sim 4.75 \text{ \AA}$).
Nevertheless, the doublet is unresolved when the velocity dispersion is large, either due to an intrinsically large dispersion or due to beam-smearing effects on galaxies. Based on the average line ratio estimated in the regions where the doublet is well resolved in the structure, we decided to constrain the ratio of \oiib\ over \oiia\ line fluxes to between 1.3 and 1.5 (see section \ref{electron_density_temperature}).

\begin{figure}
 \includegraphics[width=9cm]{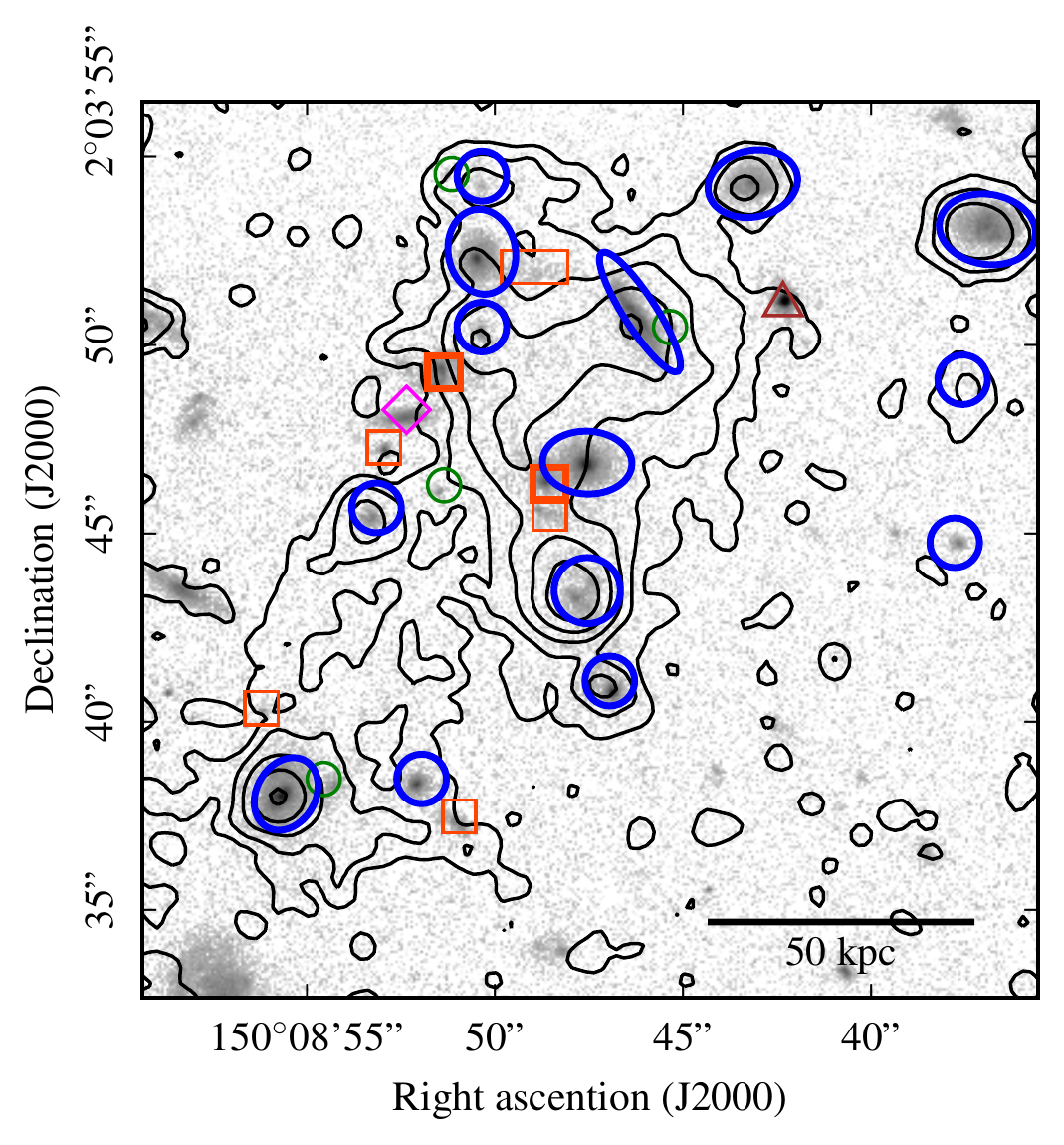}
 \caption{HST-ACS image (F814W filter, logarithmic scale, arbitrary unit) with membership identification: galaxy group members are marked using blue ellipses (which define the apertures used to extract spectra), foreground and background objects are marked using magenta diamonds and red rectangles, respectively, a star is identified by a brown triangle, and objects with undefined spectroscopic redshifts are shown with green circles. Thicker red symbols correspond to two background galaxies, which we refer to as CGr30-237 (the most central) and CGr30-86 (the northernmost) and which show \mgii\ in absorption at the redshift of the ionised structure. Logarithmic \oii\ contours, smoothed using a 0.2'' FWHM Gaussian, are overlaid in black. Contour levels correspond to 1.5, 3.3, 7.3, 16.2 and $35.9 \times 10^{-18}$~erg~s$^{-1}$~cm$^{-2}$~arcsec$^{-2}$. The physical scale at the redshift of the structure is indicated at the bottom right.
 }
 \label{hst_members}
\end{figure}

\begin{figure}
 \includegraphics[width=9cm]{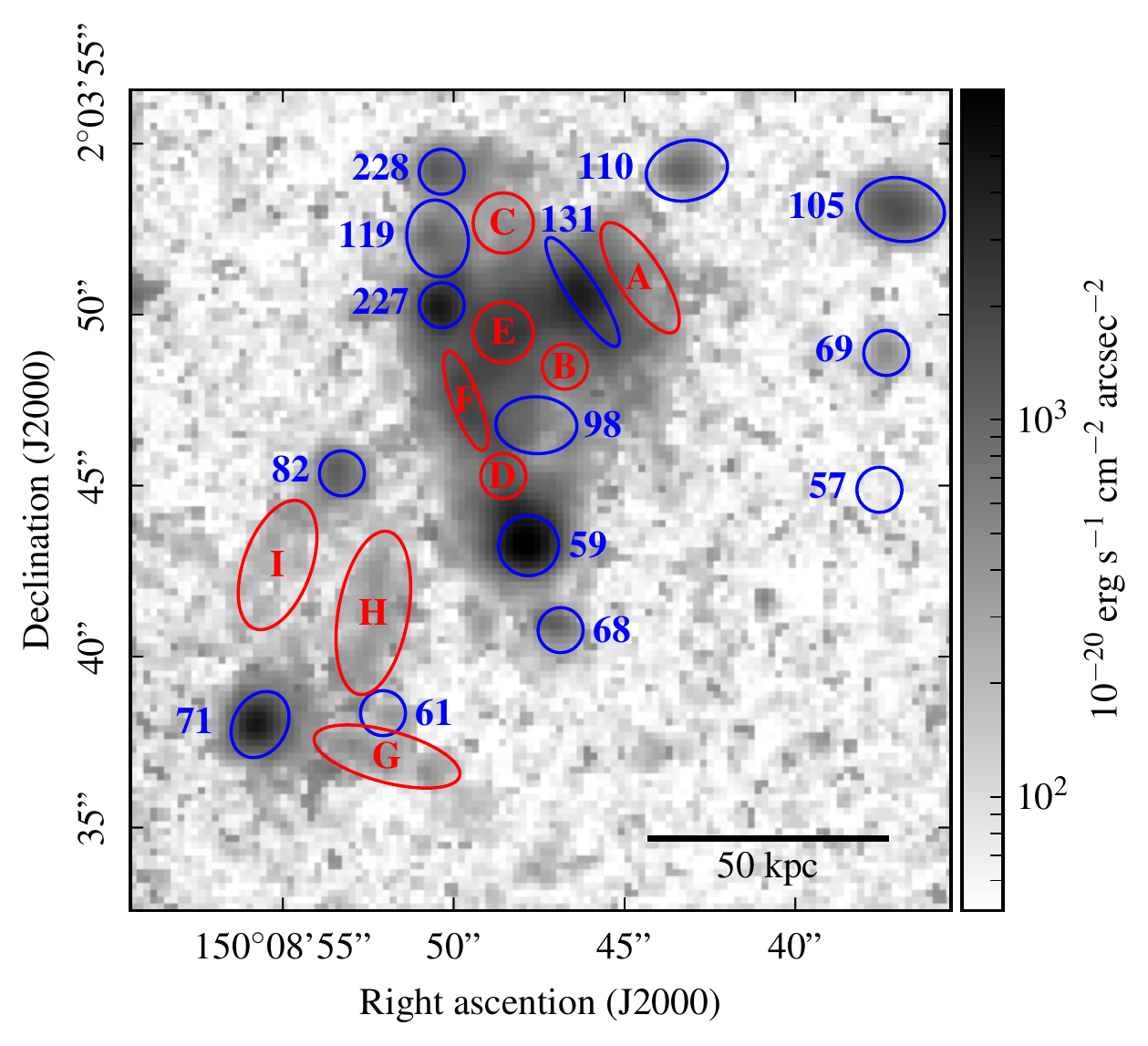}
 \caption{MUSE \oiiab\ flux map (logarithmic scale) with apertures defined to extract spectra in red for diffuse emission regions and in blue for group galaxies. The IDs of the extended gas regions and galaxies are indicated. The physical scale at the redshift of the structure is indicated at the bottom right.}
 \label{oii_ids}
\end{figure}

The \oiiab\ flux map is shown in Figure \ref{hst_members} as contours on top of the HST-ACS image in band F814W and in Figure \ref{oii_ids} as an image.
Fourteen galaxies in this area are unambiguously identified as galaxy group members from various spectral features (cf. section \ref{gal_identification}). Eleven of them have a clear detection in \oii. However, the \oii\ doublet is also detected over almost $10^4$~kpc$^2$ , and its emission extends both in between and beyond galaxies. In regions between galaxies, \oii\ emission is not uniform and can display some filaments as well as more concentrated emission. 
In these diffuse regions, the \oii\ surface brightness is measured without ambiguity down to the detection limit per spaxel, estimated to be $\sim 1.5\times 10^{-18}$~erg~s$^{-1}$~cm$^{-2}$~arcsec$^{-2}$ at 3$\sigma$,
and increases locally up to $\sim 2\times 10^{-17}$~erg~s$^{-1}$~cm$^{-2}$~arcsec$^{-2}$. The highest surface brightness regions are therefore not always centred on the galaxies involved in the structure.

\begin{figure}
 \includegraphics[width=9cm]{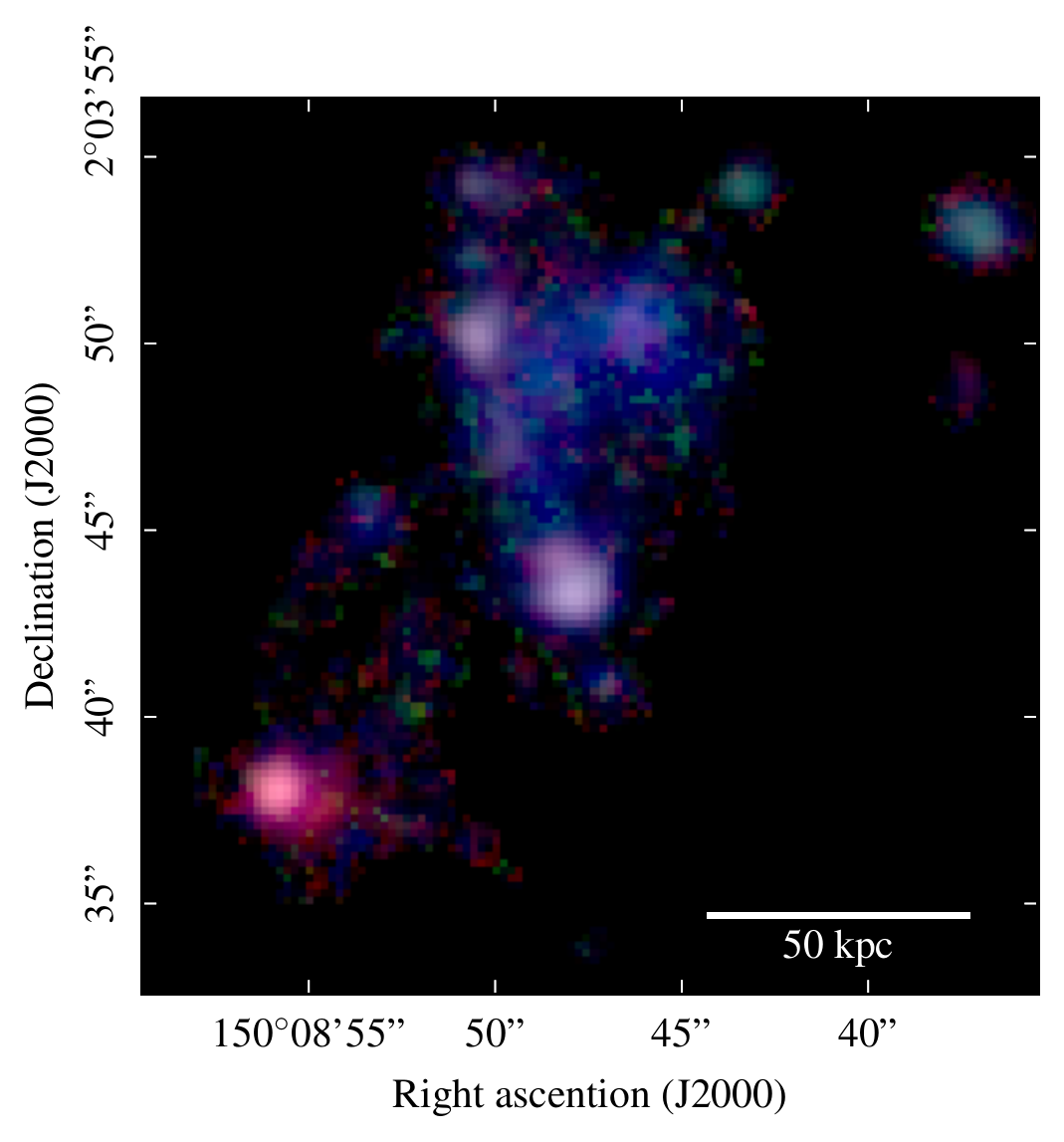}
 \caption{Flux map of the three main emission lines on a logarithmic scale: the \oiii\ doublet in red, \hb\ in green and the \oii\ doublet in blue. The \hb\ line is noisier than the two doublets. The flux maps were cleaned before display to mask regions with a too low signal-to-noise ratio using a threshold of $1.5 \times 10^{-18}$~erg~s$^{-1}$~cm$^{-2}$~arcsec$^{-2}$ on the \oii\ surface brightness map smoothed using a 0.28'' FWHM Gaussian. The physical scale at the redshift of the structure is indicated at the bottom right.
 }
 \label{oiii_hb_oii_rgb}
\end{figure}

Some \hb\ and \oiiiab\ emission is also detected in the extended gas regions where \oii\ is the brightest. On average, the flux of these lines is three to five times fainter than \oii. Flux maps of the \hb\ line and \oiiiab\ doublet have been extracted independently using the same procedure as for the \oii\ doublet described above. A composite image combining the three main emissions lines is shown in Figure \ref{oiii_hb_oii_rgb}.

\subsection{Group membership}
\label{gal_identification}

In order to determine which galaxies belong to the structure and which are foreground or background sources, we have estimated spectroscopic redshifts of the sources in the region covered by the ionised gas using their emission and absorption spectral features in MUSE spectra.
For each object of the COSMOS2015 catalogue \citep{Laigle:2016} in the considered field of view (no magnitude limit), a PSF-weighted spectrum was extracted, as was done by \citet{Inami:2017}, for
instance. We ran a customised version of the redshift-finding algorithm \emph{MARZ} \citep{Hinton:2016, Inami:2017} on these spectra.
At the redshift of the group ($z\sim 0.725$), the main emission lines that can be used to probe group membership are the \oiiab\ doublet, the \hb\ line (and any lower energy Balmer lines), and the \oiiiab\ lines. The main stellar absorption features are \ion{Ca}{ii} H$\lambda$3968.47 and \ion{Ca}{ii} K$\lambda$3933.68 lines and Balmer absorption lines.
We also visually investigated the MUSE data cube around the wavelength of the \oii\ doublet at the redshift of the group to search for additional group members that are absent from the COSMOS2015 catalogue and found two sources (CGr30-227 and CGr30-228) that both clearly have optical counterparts.
Some galaxies are spatially blended in MUSE data becauses of the seeing-limited spatial resolution, but are clearly separated in the HST image. They affect the spectra of the diffuse gas and of some group members (emission and absorption lines as well as continuum).
Using spectra together with images, we were able to distinguish distinct spectral features for these blended sources.
All galaxies with a redshift identified spectroscopically within or close to the extended structure are marked in Figure \ref{hst_members}. 
Fourteen galaxies in the considered area are unambiguously identified as galaxy group members. They all show some weak or strong emission lines, except for one galaxy, CGr30-57.
The two thicker red squares identify two background galaxies that show absorption lines at the redshift of the extended ionised gas.

\subsection{Properties of group galaxies}
\label{galprop}

The comparison of physical properties in extended gas regions and in group members embedded in the structure is useful to understand the origin of the extended gas and its ionisation source. Studying group member morphologies and global properties provides information on the interplay between galaxies and extended gas.

\subsubsection{Galaxies morphology}
\label{morphology}

High-resolution morphology was derived for group galaxies in order to constrain the projection parameters of galaxy discs on sky (inclination and position angle of major axis) as well as the bulge and disc sizes and their relative strength.
This analysis is necessary to gather information on the geometry of the structure, but also to extract galaxy spectra.

We used HST-ACS images obtained in the F814W filter since this dataset has the best spatial resolution among the ancillary data.
These images were produced using the MultiDrizzle softwares \citep{Koekemoer:2007} on the COSMOS field \citep{Scoville:2007}.
They have a scale of 0.03''/pixel and a median exposure of 2028 seconds.

We followed the same method as in \citet{Contini:2016} for galaxies in the Hubble Deep Field South (HDFS). We used \textsc{Galfit} \citep{Peng:2002} to model the galaxy morphology.
First, we identified nine unsaturated stars over four groups that we observed with MUSE in the COSMOS field in order to determine the point-spread function (PSF) in the HST-ACS images. The usage of several groups was necessary to have enough stars. These stars were modelled using a circular Moffat profile. The theoretical PSF we used to extract the morphology was built from the median value of each parameter of all the nine stars: $FWHM=0.084$'' and $\beta=1.9$ (Moffat index).
Galaxies were modelled using a disc and a bulge. The bulge was spherical and described by a classical de Vaucouleurs profile, whereas the disc was described by an exponential disc. The free parameters were the bulge central brightness and effective radius, the disc central brightness, axis ratio, position angle of the major axis and effective radius and the centre, which the disc and the bulge share.
When necessary (e.g. for blended sources), we used additional components to account for extra features and avoid the bias they might induce on galaxy parameters.

Morphological parameters are provided in Table \ref{morph_glob_table}. We stress that the disc effective radius of CGr30-59 is clearly overestimated. This galaxy shows an asymmetric light distribution with some diffuse extent on one side.
This perturbed morphology, which is probably due to some interaction, cannot be accounted for with our simple disc plus bulge model, and we were not able to satisfactorily model this diffuse extent.
The morphology analysis was not performed for three galaxies, CGr30-69, CGr30-227 and CGr30-228, because they are too small and too faint in the HST images to obtain reliable parameters.

\subsubsection{Global properties}
\label{global}

The stellar masses, star formation rates (SFR), and extinctions of the galaxies embedded in the ionised structure were estimated using extensive photometry available in the COSMOS field \citep{Scoville:2007} described in section \ref{ancillary}.
We used photometric measurements over 3'' apertures in 32 bands from the COSMOS2015 catalogue \citep{Laigle:2016} to constrain stellar population synthesis (SPS) models.
We decided not to use the parameters produced by \citet{Laigle:2016} because this catalogue is purely photometric, whereas we have robust measurements of spectroscopic redshifts that we can use as constraints in SPS models. We used the FAST spectral energy distribution (SED) fitting code \citep{Kriek:2009} with a library of synthetic spectra generated with the SPS model of \citet{Conroy:2010}, assuming a \citet{Chabrier:2003} initial mass function (IMF), an exponentially declining SFR ($\text{SFR} \propto \exp{(-t/\tau)}$, with $8.5<\log{(\tau [\text{yr}^{-1}])}<10$), and a \citet{Calzetti:2000} extinction law.
We added in quadrature a 0.05 dex uncertainty to each band in order to derive formal uncertainties on the SED measurements
that account for residual calibration uncertainties. In general, systematic uncertainties due to different modelling assumptions can amount to $\sim 0.2$ dex in stellar mass and $\sim 0.3$ dex in SFR \citep[e.g.][and references therein]{Conroy:2013}, which is the main source of discrepancy when comparing to other SED fitting methods or to other datasets.
For the galaxies with a good match between photometric and spectroscopic redshifts, we checked that the COSMOS2015 mass estimates were compatible with the estimates used here and found a good agreement.
We note that the fit for CGr30-71 leads to a high $\chi^2$ value. The reason may be that this galaxy contains an AGN (see section \ref{ionised_measurements}) and we used no AGN template.

As mentioned above, some galaxies at different redshifts are blended in seeing-limited images and in MUSE spectra (cf. section \ref{gal_identification} and Figure \ref{hst_members}). For the group galaxy CGr30-98 most specifically, since object extraction was made using a combination of seeing-limited images in near-infrared and z-bands, the global properties were determined using a 3'' aperture that also covered a $z=0.938$ galaxy (CGr30-237). Interestingly, the photometric analysis using FAST with no constraint on the redshift led to a redshift consistent with the group. We have derived a correction for the mass of CGr30-98 using the HST-ACS image, which is the only one where the two sources can be clearly separated. We assumed that the F814W band is representative of the mass of the two sources so that the fraction of flux in each galaxy corresponds to their fraction of mass. Defining one aperture on each object and similar apertures on sky to remove background, we found that the flux in CGr30-98 is three times higher than the flux in the $z=0.938$ galaxy. Hence the correction factor for its mass is estimated to 0.75 and the mass of CGr30-98 is M$_* = 10^{10.70}$~M$_\odot$. It remains one of the two most massive galaxies in the structure.

Another consequence of the extraction using near-infrared and z-bands images is that faint and blue objects were not included in the COSMOS2015 catalogue. This is the case for two galaxies in the structure, CGr30-227 and CGr30-228 (cf. Figure \ref{oii_ids}). These two galaxies are not part of previous COSMOS catalogues \citep{Capak:2007,Ilbert:2009} either. A full extraction of photometry would be needed, however, we are mainly interested in knowing the mass of the most massive galaxies, and it is sufficient to estimate an upper limit for these faint galaxies.

Parameters (stellar mass, SFR, and extinction) extracted from the SED fitting are given in Table \ref{morph_glob_table}.

\begin{table*}[t]
\caption{Morphological and global parameters of
the group galaxies.
Column 1: MUSE ID for the COSMOS-Group 30. Column 2: COSMOS2015 ID except for CGr30-227 and CGr30-228, which are not in the catalogue. Column 3: MUSE spectroscopic redshift. Columns 4-6: Extinction, stellar mass, and star formation rate from the SED fitting code FAST.
Columns 7-9: Exponential disc inclination, effective radius, and position angle of the major axis derived using \textsc{Galfit}. The asterisk$^*$ indicates that the radius is not well constrained for the faint galaxy CGr30-59, which has unusual features.}
\begin{center}
 \begin{tabular}{ccccccccc}
\hline
CGr30-ID & COSMOS2015 ID & Redshift & $E(B-V)$ [mag] & $\log{\frac{\text{M}_*}{\text{M}_\odot}}$ & $\log{\frac{\text{SFR}}{\text{M}_\odot \text{yr}^{-1}}}$ & i [$^\circ$] & $r_e$ [kpc] & PA [$^\circ$] \\
 (1) & (2) & (3) & (4)  & (5) & (6)  & (7)  & (8) & (9) \\
\hline
57 & 501021 & 0.7251 & $0.00^{+0.01}$ & $9.47^{+0.01}_{-0.28}$ & $-4.45^{+2.88}_{-3.00}$ & 45 & 1.5 & 72 \\
59 & 501090 & 0.7232 & $0.22^{+0.01}_{-0.00}$ & $9.11^{+0.03}_{-0.01}$ & $1.00^{+0.01}_{-0.03}$ & 74 & 10.9$^*$ & 16 \\
61 & 501214 & 0.7257 & $0.62^{+0.01}_{-0.10}$ & $9.78^{+0.06}_{-0.03}$ & $1.79^{+0.04}_{-0.42}$ & 80 & 1.5 & -51 \\
68 & 501562 & 0.7242 & $0.20^{+0.17}_{-0.07}$ & $10.18^{+0.24}_{-0.03}$ & $0.13^{+0.38}_{-0.12}$ & 88 & 0.3 & -4 \\
69 & 501572 & 0.7256 & $0.15^{+0.34}_{-0.15}$ & $8.41^{+0.32}_{-0.38}$ & $-1.27^{+1.54}_{-0.88}$ & - & - & - \\
71 & 501652 & 0.7248 & $0.62^{+0.03}_{-0.05}$ & $10.90^{+0.08}_{-0.01}$ & $2.05^{+0.10}_{-0.11}$ & 40 & 3.4 & -30 \\
82 & 502157 & 0.7267 & $0.57^{+0.11}_{-0.22}$ & $9.56^{+0.35}_{-0.10}$ & $1.26^{+0.29}_{-0.65}$ & 26 & 1.7 & -75 \\
98 & 503138 & 0.7243 & $0.02^{+0.03}_{-0.02}$ & $10.70^{+0.13}_{-0.09}$ & $-0.05^{+0.11}_{-0.05}$ & 46 & 3.9 & 88 \\
105 & 503407 & 0.7267 & $0.42^{+0.00}_{-0.15}$ & $10.49^{+0.01}_{-0.10}$ & $0.82^{+0.01}_{-0.13}$ & 44 & 4.2 & 81 \\
110 & 503517 & 0.7268 & $0.32^{+0.02}_{-0.13}$ & $10.08^{+0.23}_{-0.08}$ & $0.40^{+0.26}_{-0.20}$ & 43 & 4.0 & -78 \\
119 & 504056 & 0.7242 & $0.30^{+0.10}_{-0.12}$ & $10.53^{+0.10}_{-0.22}$ & $0.64^{+0.34}_{-0.31}$ & 38 & 3.7 & 10 \\
131 & 504620 & 0.7232 & $0.40^{+0.05}_{-0.05}$ & $10.48^{+0.01}_{-0.04}$ & $1.34^{+0.13}_{-0.20}$ & 77 & 6.2 & 33 \\
227 & - & 0.7239 & - & - & - & - & - & - \\
228 & - & 0.7243 & - & - & - & - & - & - \\
\hline
 \end{tabular}
\end{center}
\label{morph_glob_table}
\end{table*}

\subsection{MUSE spectra analysis}

MUSE spectra contain much information for sources at $z\sim 0.7$, most specifically, on the ionised gas properties. Comparing properties of this gas in extended diffuse regions and in galaxies embedded in the structure will give insights on the origin of the extended gas and on the possible ionisation source.
We exclude CGr30-57 from the subsequent analysis because it has no emission line associated with ionised gas, as expected from its very low SFR.

\subsubsection{Spectra extraction}
\label{apertures}

On the one hand, we visually defined various apertures on regions with extended emission, but without any counterpart in broad-band images. Their position and size were adjusted individually based on kinematics and \oii\ flux map features, such as filaments or bright regions with coherent velocities, avoiding \oii\ emission from galaxies. 
These regions are shown in Figure \ref{oii_ids} as red ellipses with their corresponding ID.

On the other hand, for each galaxy belonging to the group, an elliptical aperture was defined from the morphological analysis performed on the HST-ACS image (section \ref{morphology}). The shape and orientation of the ellipse were set to that of the disc: major axis orientation, axis ratio, and size equal to $2.2\times r_e$ , where $r_e$ is the disc effective radius. For a purely exponential disc, $\sim 90$\% of the flux is contained inside $2.2\times r_e$.  Therefore using this radius allows us to have most of the flux from the optical part of the galaxy without probing the region where extended emission may dominate. For the same reason, the shape of the aperture was not corrected for the PSF, which would have enlarged the aperture.
When galaxies were either smaller than the spatial resolution of the MUSE data cube ($2.2 \times r_e < 0.68$'') or barely detected in HST images, we defined circular apertures with a diameter of 3 pixels corresponding to 0.6'', except for CGr30-59, for which we defined an aperture of 4 pixels (0.8'') because of the large extent of the \oii\ emission. 
These apertures are shown in Figures \ref{hst_members} and \ref{oii_ids} with blue ellipses.

For each aperture, an integrated spectrum was obtained by adding the spectrum of all the spaxels within the corresponding aperture. An integrated variance spectrum was obtained in the same way.

\subsubsection{Continuum removal}
\label{continuum_removal}

The continuum of galaxies is commonly affected by
absorption lines that lie below or close to emission lines. Spectra of extended gas regions do not show evidence of continuum emission. However, in some cases, continuum from background or foreground sources can contribute to the spectra (e.g. CGr30-D, CGr30-F, and CGr30-G).
In order to study emission line properties with a minimum contamination from continuum features, we used the penalised pixel-fitting (\emph{pPXF}) code\footnote{\url{http://www-astro.physics.ox.ac.uk/~mxc/software/}} \citep{Cappellari:2004} to adjust the stellar continuum in the spectra of both galaxies and extended gas regions. 
The MILES stellar population synthesis library (spectra from 3525 - 7500 \AA\ at a spectral resolution of 2.5 \AA\ FWHM, \citealp{Sanchez-Blazquez:2006,Vazdekis:2010,Falcon-Barroso:2011}) was used although its resolution is lower than that of MUSE for galaxies at $z\sim 0.7$.
However, we are interested in removing stellar continuum contribution from the spectrum in order to perform accurate emission line measurements, in particular around the Balmer line series, where significant absorption can affect line measurements. Recovering the stellar velocity dispersion is therefore not our goal here.
The observed velocity dispersion in the integrated spectra is the combination of intrinsic dispersion, large-scale motions, and the MUSE LSF. It is higher than the MILES resolution for galaxies with a significant continuum, which are the largest and most massive ones (see Figures \ref{vfs} and \ref{stellarkin_maps}). The MILES library therefore enables modelling the spectra accurately, and it is well suited to recover the spectrum of $z\sim 0.725$ galaxies in the optical thanks to its wavelength coverage. 

We used a subset of 156 SED template spectra \citep{Vazdekis:2010} with a default linear sampling of 0.9 \AA/pix and a default spectral resolution of $\text{FWHM}=2.51$ \AA. We used templates generated using an unimodal Salpeter IMF with a slope of 1.30 and selected a range of population parameters over a regular grid of six steps in $[M/H]$ from -1.71 to 0.22 and of 26 logarithmic steps in age from 1 to 16 Gyr. Even though stellar populations older than 7 Gyr may not be realistic for $z\sim 0.725$ galaxies, the adopted templates are sufficient for a good fit of the stellar absorption lines, which is our goal. We also checked that our galaxies were not dominated by stellar populations older than 7 Gyr. This was
the case for only three galaxies, which do not exhibit strong enough emission lines to make a more robust analysis useful. In most of the cases, the bulk of stellar populations is young ($\sim 1$ Gyr).
In addition to this stellar library, some emission lines were included in the templates, and the stellar continuum template was multiplied by a Legendre polynomial of degree 10 to account for possible calibration uncertainties (both in templates and MUSE data).

When \emph{pPXF} converged to the best solution, we generated a continuum-only spectrum from the model that we removed from the observed spectrum to generate a continuum-free spectrum.

\subsubsection{Emission line measurements}
\label{ionised_measurements}

Emission line fluxes were computed using a modified version of \emph{CAMEL} for 1D spectra\footnote{called \emph{SAMEL} in the \emph{CAMEL} package} and fitting all lines available in the continuum-free spectra simultaneously. The choice of using \emph{CAMEL} rather than \emph{pPXF} for emission line measurements was motivated by (i) the fact that \emph{CAMEL} fits emission lines with a linear sampling instead of a logarithmic sampling in \emph{pPXF}, that is, without any interpolation of the data, and (ii) the
better control we had on trying to place some constraints on the fits (e.g. line ratio).
Finally, emission lines were fitted without any constraint on line ratio because the extinction is not known a priori and because the \oii\ doublet line ratio can be unusual (see section \ref{electron_density_temperature}).

\begin{figure*}
\includegraphics[width=18cm]{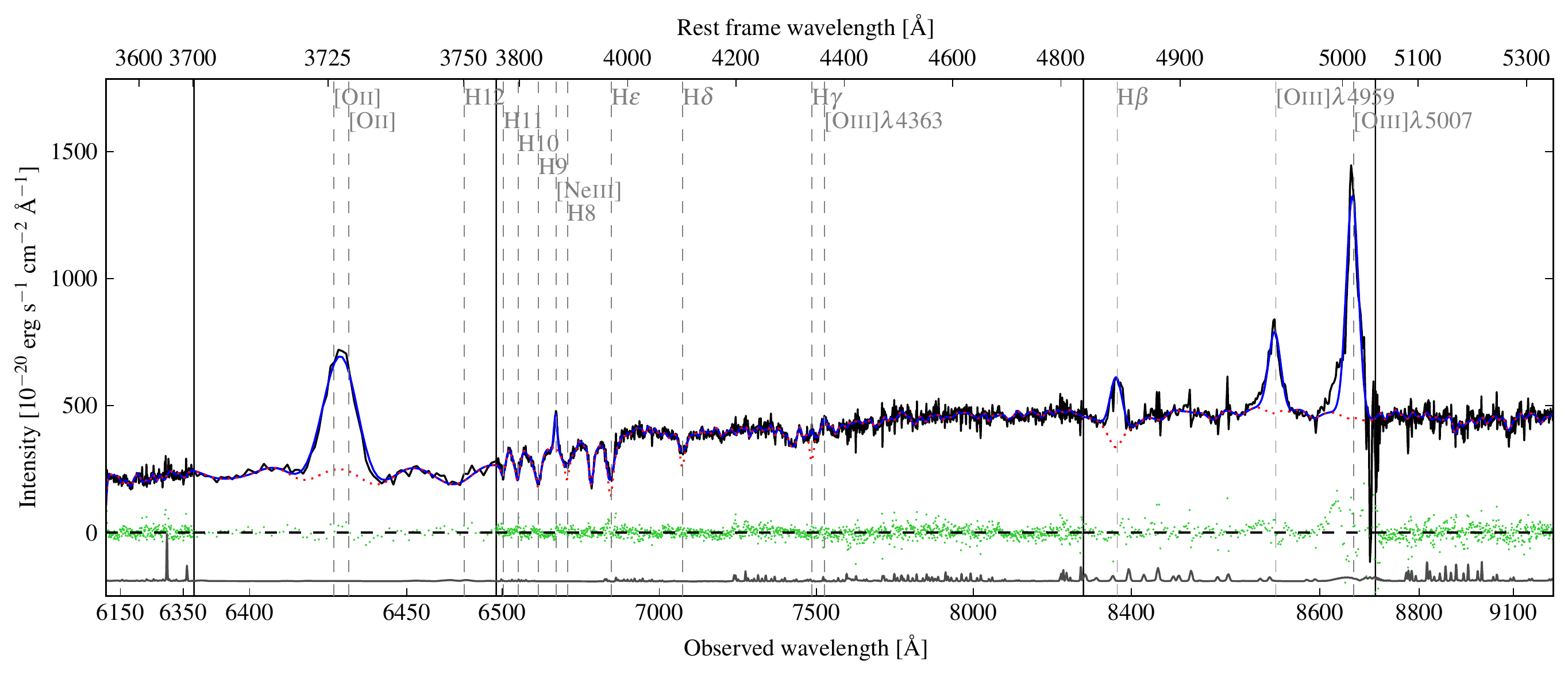}
\includegraphics[width=18cm]{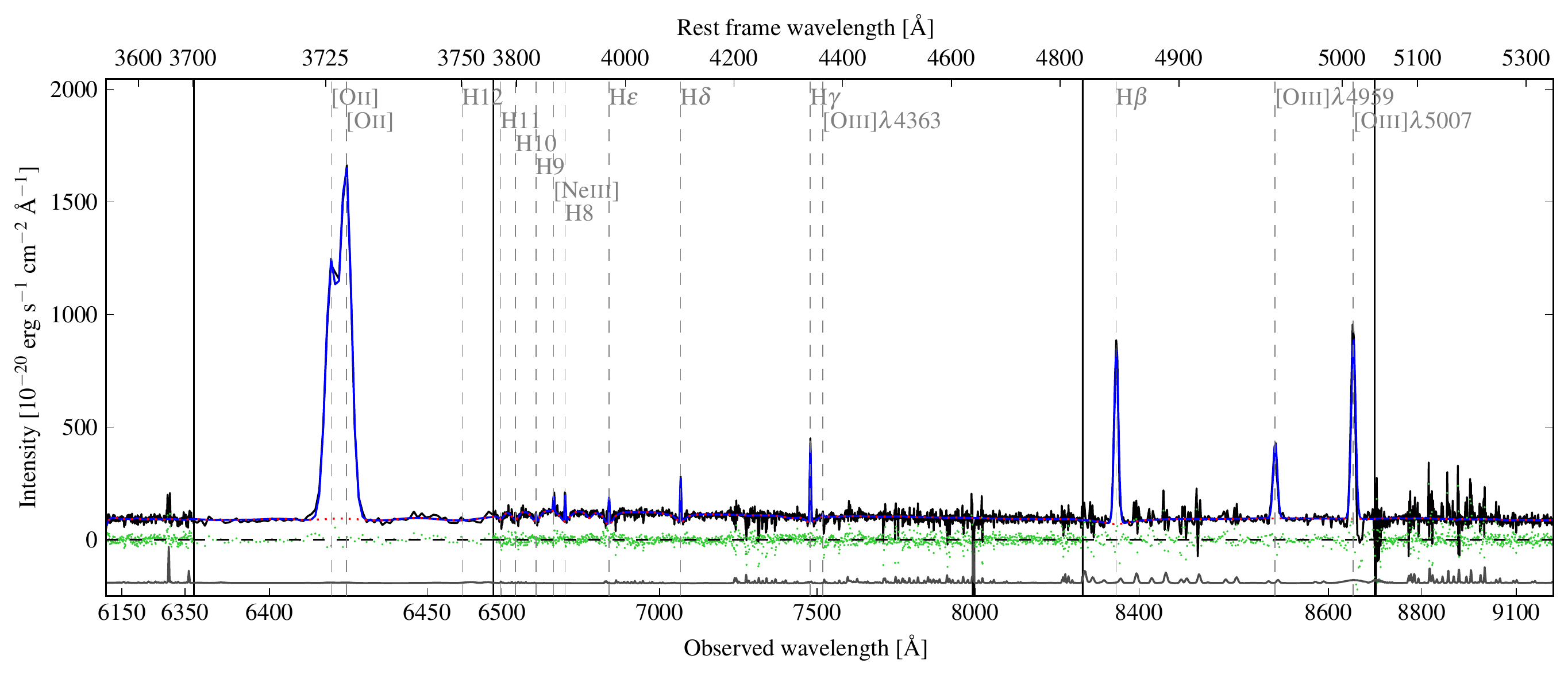}
 \caption{Examples of spectra for CGr30-71 (top) and CGr30-59 (bottom). In each panel, the MUSE integrated spectrum (black) is plotted with \emph{pPXF} continumm (dotted red) plus \emph{CAMEL} emission lines model (blue), model residuals (green dots), and the variance spectrum (grey), which is shifted below zero for clarity. The name and position of the main emission lines are indicated in grey. Black vertical lines separate spectral ranges displayed with different scales to zoom in on the main emission lines. In the top panel, the broadening of emission lines due to the AGN is clearly visible. In the bottom panel, the \oii\ doublet is well resolved.
}
 \label{spectra_ex}
\end{figure*}

Figure \ref{spectra_ex} shows two examples of observed and modelled spectra for CGr30-71 (top) and CGr30-59 (bottom).
Residuals are displayed together with the variance spectrum. For CGr30-71, absorption features are clear, and the \hb\ emission line is affected by an absorption component, whereas the spectrum of CGr30-59 has a weaker continuum, but clearly displays several Balmer lines and the resolved \oii\ doublet.

Fluxes of the main emission lines are listed in Table \ref{lines_table}. They are corrected for Galactic extinction. This correction was performed using the \citet{Schlegel:1998} measurements in the direction of the ionised gas structure. On average over the considered sky area, the reddening is estimated to be $E(B-V) = 0.0178$\footnote{\url{http://irsa.ipac.caltech.edu/applications/DUST/}}. The Galactic extinction was taken into account using the Milky Way attenuation curve from \citet{Cardelli:1989} and assuming a classical value of 3.1 for the ratio between visual extinction and reddening $R_V= A_V / E(B-V)$. When a line was not detected, a detection threshold was computed for the corresponding aperture. The 1$\sigma$ threshold was estimated as the flux of a Gaussian line assuming the same dispersion as for detected lines and with an intensity equal to the standard deviation of the residual spectrum (weighted by variance) within $\text{approximately
five}$ times the dispersion of the line around the expected line position. In some case, these thresholds can differ significantly from (1$\sigma$) uncertainties inferred from the fits. These variations can be due to the methods that are different, but also to the various sizes of the apertures and their specific location in the data cube (both spatially and spectrally).

\begin{table*}[t]
\caption{
Emission-lines fluxes in group galaxies and extended gas regions.
Column 1: MUSE ID for the COSMOS-Group 30. Alphabetic IDs correspond to apertures on diffuse gas, numeric IDs correspond to galaxies. Column 2: Area of each aperture in kpc$^2$. Columns 3-7: Main emission line integrated flux corrected for Galactic extinction in units of $10^{-18}$ erg s$^{-1}$~cm$^{-2}$. When the \oii\ doublet is resolved, the fluxes of both \oiia\ and \oiib\ are provided, otherwise their sum is indicated. When a line is not detected, the 3$\sigma$ detection threshold is given. Column
7: Other emission lines identified above the 3$\sigma$ detection threshold are provided when they are not embedded in strong absorption lines.}
\begin{center}
 \begin{tabular}{cccccccccc}
\hline
CGr30-ID & Area [kpc$^2$] & \oiiab & \oiiib & \oiiia & \hb & \hg & Other ELs \\
(1) & (2) & (3) & (4) & (5) & (6) & (7) & (8) \\
\hline
A        &   204 & $20.8 \pm 5.1$                     & $<6.9$          & $<12.7$         & $<9.9$          & $<6.4$          & - \\
B        &    78 & $13.9 \pm 5.9$                     & $<4.5$          & $<6.3$          & $<6.2$          & $<3.4$          & - \\
C        &   128 & $9.5 \pm 1.3$                      & $<3.2$          & $<6.4$          & $<5.0$          & $<2.7$          & - \\ 
D        &    78 & $19.6 \pm 1.9$                     & $<2.5$          & $3.3 \pm 1.7$   & $4.3 \pm 0.8$   & $<1.3$          & - \\ 
E        &   128 & $54.1 \pm 4.1$                     & $<4.4$          & $<8.1$          & $11.6 \pm 0.8$  & $7.0 \pm 0.6$   & \hd \\ 
F        &    96 & $11.3 \pm 0.7$ | $21.2 \pm 0.6$    & $2.4 \pm 0.6$   & $7.5 \pm 1.0$   & $10.1 \pm 0.8$  & $5.6 \pm 0.6$   & \hd \\ 
G        &   275 & $15.3 \pm 1.7$                     & $<5.7$          & $9.3 \pm 2.0$   & $<6.3$          & $<3.5$          & - \\ 
H        &   410 & $21.6 \pm 1.8$                     & $<7.3$          & $<19.2$         & $<11.3$         & $<8.7$          & - \\
I        &   322 & $13.2 \pm 2.6$                     & $<5.5$          & $<9.0$          & $<11.6$         & $<5.2$          & - \\
\hline
59       &   130 & $50.1 \pm 0.9$ | $70.2 \pm 0.9$    & $19.8 \pm 0.8$  & $48.1 \pm 1.8$  & $44.8 \pm 0.9$  & $18.0 \pm 1.0$  & \neiiia, \hd\ to H9 \\ 
61       &    67 & $4.1 \pm 2.4$                      & $<5.9$          & $<7.8$          & $<8.5$          & $<5.1$          & - \\
68       &    67 & $7.8 \pm 1.0$                      & $<2.6$          & $<5.8$          & $<4.3$          & $<2.3$          & - \\  
69       &    78 & $2.0 \pm 0.2$ | $2.5 \pm 0.2$      & $<1.2$          & $2.6 \pm 0.4$   & $<1.4$          & $<0.9$          & - \\
71       &   130 & $62.3 \pm 3.6$                     & $52.9 \pm 1.8$  & $146.1 \pm 3.5$ & $44.8 \pm 1.9$  & $17.9 \pm 1.5$  & \neva, \neiiia, \hd \\ 
82       &    67 & $11.7 \pm 1.3$                     & $<3.6$          & $<6.0$          & $4.5 \pm 0.5$   & $<2.6$          & - \\ 
98       &   162 & $29.8 \pm 5.8$                     & $<8.1$          & $<13.0$         & $<14.1$         & $<6.9$          & - \\
105      &   198 & $30.7 \pm 1.3$                     & $<2.7$          & $6.7 \pm 1.1$   & $26.8 \pm 0.9$  & $13.2 \pm 0.7$  & \hd \\ 
110      &   175 & $5.2 \pm 0.4$ | $7.3 \pm 0.4$      & $<1.3$          & $<2.0$          & $15.4 \pm 0.5$  & $6.9 \pm 0.4$   & \hd \\
119      &   164 & $8.7 \pm 0.6$ | $11.3 \pm 0.5$     & $<2.8$          & $5.0 \pm 1.2$   & $12.5 \pm 1.2$  & $4.6 \pm 0.7$   & - \\ 
131      &   132 & $67.8 \pm 6.6$                     & $<7.6$          & $13.5 \pm 1.7$  & $18.7 \pm 0.9$  & $9.8 \pm 1.0$   & - \\ 
227      &    78 & $19.3 \pm 0.4$ | $29.0 \pm 0.5$    & $8.5 \pm 0.8$   & $21.8 \pm 0.9$  & $19.0 \pm 0.8$  & $8.1 \pm 0.6$   & \neiiia, \hd\ to H9 \\ 
228      &    71 & $5.3 \pm 0.3$ | $7.7 \pm 0.3$      & $1.5 \pm 0.4$   & $5.6 \pm 0.6$   & $5.4 \pm 0.6$   & $<1.2$          & - \\ 
\hline
 \end{tabular}
\end{center}
\label{lines_table}
\end{table*}

\subsection{Kinematics}

The kinematics of the extended gas region can provide some indications on the nature of this structure. In addition, comparing these kinematics with both gaseous and stellar kinematics of group galaxies, when possible, may help understanding the interplay between galaxies and diffuse gas.

\subsubsection{Ionised gas kinematics}
\label{gaskin}

\begin{figure}
 \includegraphics[width=9cm]{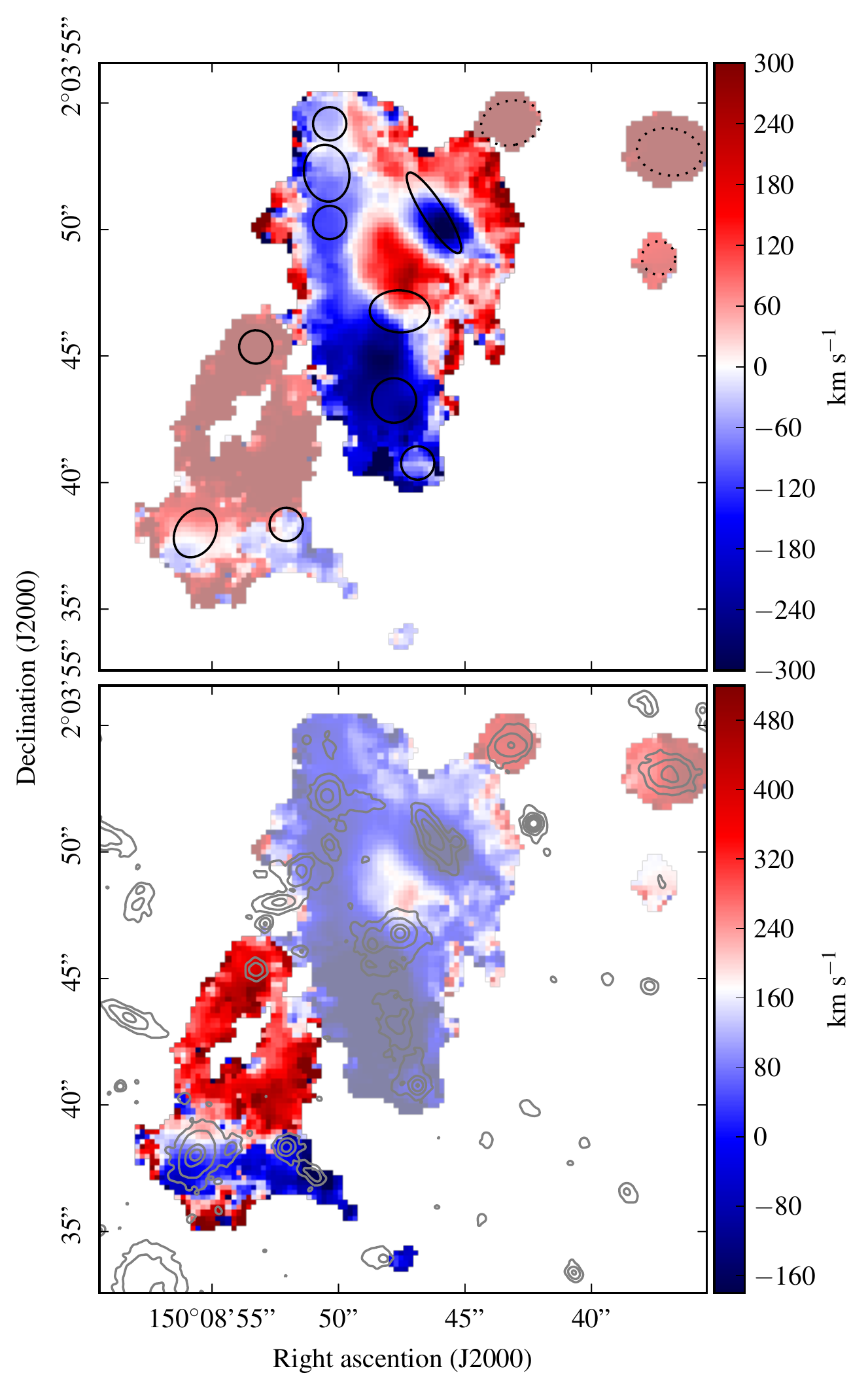}
 \caption{\oii\ velocity fields over a 25'' square field of view using two different velocity ranges centred on the northern (top) and southern (bottom) components highlighted in each panel. A threshold of $1.5 \times 10^{-18}$~erg~s$^{-1}$~cm$^{-2}$~arcsec$^{-2}$ has been applied on the \oii\ surface brightness map smoothed using a 0.28'' FWHM Gaussian, to mask regions with a too low signal-to-noise ratio. Regions outside each component are faded to distinguish them better. In the top panel, group members are overlaid using black ellipses. Three galaxies do not seem directly linked to any extended gas region and are indicated with dotted ellipses. In the bottom panel, HST contours are overlaid in grey. They are logarithmic and smoothed using a 0.09'' FWHM Gaussian. 
 }
 \label{vfs}
\end{figure}

The \oii\ velocity field generated by \emph{CAMEL} is displayed in Figure \ref{vfs} with two different velocity ranges centred on two distinct sub-structures. It displays some discontinuity that likely separates two main regions with a velocity offset larger than $\sim200$~km~s$^{-1}$ (up to $>500$~km~s$^{-1}$). The most extended ionised gas region is located on the northern side and has a smooth velocity field. It is spread over $\sim 5000$~kpc$^2$. The other region covers $\sim 2300$~kpc$^2$ (cf. section \ref{mass_mgii}) and has a filamentary structure. We refer to these two regions as northern and southern components.
In order to better follow the distribution of the ionised gas and its kinematics with respect to group members, black ellipses (top) and HST imaging contours (bottom) are overlaid on these maps.
The position and global velocity of three galaxies (CGr30-69, CGr30-105 and CGr30-110) suggest that they are not linked to the extended gas structure. They are identified with dotted ellipses in the top panel.

\begin{figure}
 \includegraphics[width=9cm]{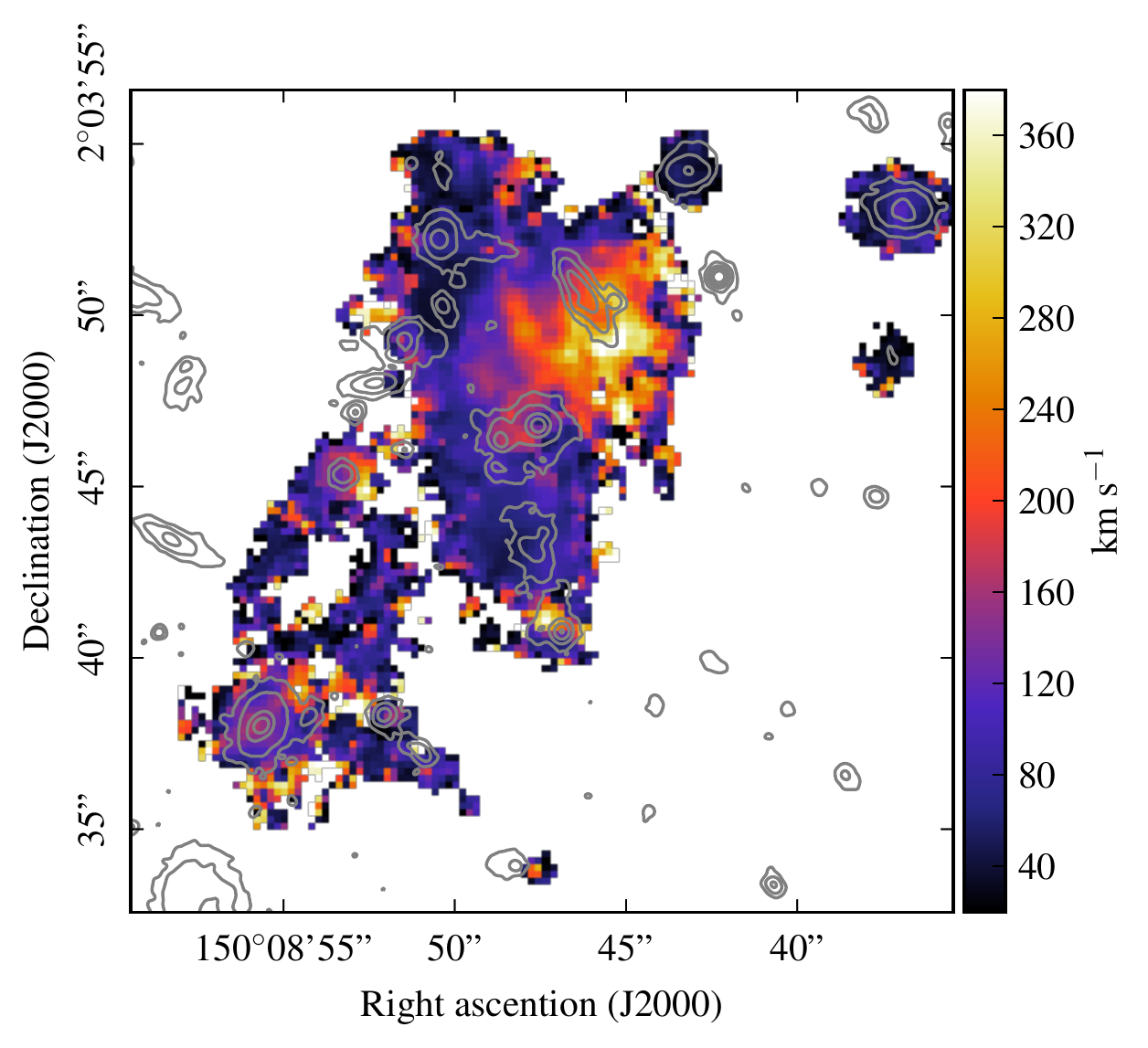}
 \caption{Velocity dispersion map corrected for the instrumental LSF
 with HST contours overlaid in grey. Contours are logarithmic and smoothed using a 0.09'' FWHM Gaussian. A threshold of $1.5 \times 10^{-18}$~erg~s$^{-1}$~cm$^{-2}$~arcsec$^{-2}$ has been applied on the \oii\ surface brightness map smoothed using a 0.28'' FWHM Gaussian, to mask regions with a too low signal-to-noise ratio.
 }
 \label{sig}
\end{figure}

The velocity dispersion map is shown in Figure \ref{sig} with HST imaging contours. It has been corrected for the instrumental LSF at the wavelength of \oii\ assuming the LSF is Gaussian (eq. \ref{lsf}).
In order to quantify the velocity dispersion for each galaxy and for each extended gas region, we computed the velocity dispersion on each aperture defined in section \ref{apertures} as the median of the velocity dispersion map. These estimates are given in Table \ref{ext_sfr_mass_table} where the associated uncertainties were estimated as the standard deviation over the regions.
The average (median) velocity dispersion in the diffuse regions is $126(99)\pm54$~km~s$^{-1}$, which is only marginally higher than in galaxies ($110(83)\pm50$ km~s$^{-1}$).
Nevertheless, velocity dispersions as low as 70~km~s$^{-1}$ are only observed in some small galaxies, but never in extended gas regions.
The velocity dispersion is above 300~km~s$^{-1}$ on and around the west side of the edge-on galaxy CGr30-131. We performed a two-component decomposition to investigate whether this broadening might be due to two components with different velocities (approaching side of this galaxy with $v\sim -300$~km~s$^{-1}$ on the one hand, diffuse gas at $v\sim 100$~km~s$^{-1}$ on the other hand).
However, when we used two components with various constraints (e.g. velocity separation, line intensities, dispersion), the velocity dispersion remained quite high. It therefore seems to be a real feature rather than an artefact due to beam smearing.

\subsubsection{Stellar kinematics of massive group galaxies}
\label{stellarkin}

The ionised gas structure seems to be connected to the most massive galaxies of this region. Therefore, studying stellar kinematics of these galaxies can give some insights on the relation of the diffuse ionised gas kinematics to stellar kinematics of massive galaxies.
This analysis can be performed for the two most massive group galaxies located in the structure, CGr30-98 and CGr30-71, thanks to their strong stellar spectra.
We have used the latest version of \emph{pPXF} \citep{Cappellari:2004, Cappellari:2017} to extract these kinematics. The method is the same as used on MUSE data of intermediate-redshift galaxies observed in the HDFS and in the Hubble Ultra Deep Field, and described in detail in \citet{Guerou:2017}. We first extracted two sub-cubes centred on each galaxy. Then, in order to remove noise-dominated pixels in the subsequent spatial binning step, we masked spectra with a signal-to-noise ratio lower than unity, estimated on the stellar continuum between 4150\,--\,4350\,\AA\, (in the galaxy rest-frame).
For the galaxy CGr30-98, it was necessary to mask the region where the $z=0.938$ background galaxy contributed to the spectra to avoid complex fitting. We then spatially binned the selected spectra to a target signal-to-noise ratio of 10 using the adaptive spatial binning software developed by \cite{Cappellari:2003} and fit the two first orders of the line-of-sight velocity distribution, namely the radial velocity, $V$, and the velocity dispersion, $\sigma$, of the stellar and gas components simultaneously. The kinematics of each component (stars and gas) was let free to vary.
To fit the stellar continuum of the spatially binned MUSE spectra, we used a subset of 53 stellar templates from the Indo-US library \citep{Valdes:2004} selected as in \citet{Shetty:2015}. The resolution of this library is FWHM $\sim$ 1.35 \AA\ \citep{Beifiori:2011}, which is well suited to study kinematics of $z\sim 0.725$ galaxies from MUSE data because it is lower than the MUSE rest-frame LSF FWHM ($\sim 1.45$ \AA) in the considered wavelength range. Gaussian emission line templates were used to describe the gas components. Both stellar and gas templates were broadened to the MUSE LSF \citep[see][]{Bacon:2017,Guerou:2017} before the fitting procedure.
We set up \emph{pPXF} to use additive polynomials of the sixth order and multiplicative polynomials of the first order, and fitted the MUSE spectra over the wavelength range 3740\,--\,5100\,\AA\,(in the object rest-frame). Finally, we first determined the best combination of stellar templates for each object by taking the best-fit solution of the galaxy stacked spectrum, and used this best template to fit each individual spatially binned spaxel. 

\begin{figure}
 \includegraphics[width=4.45cm]{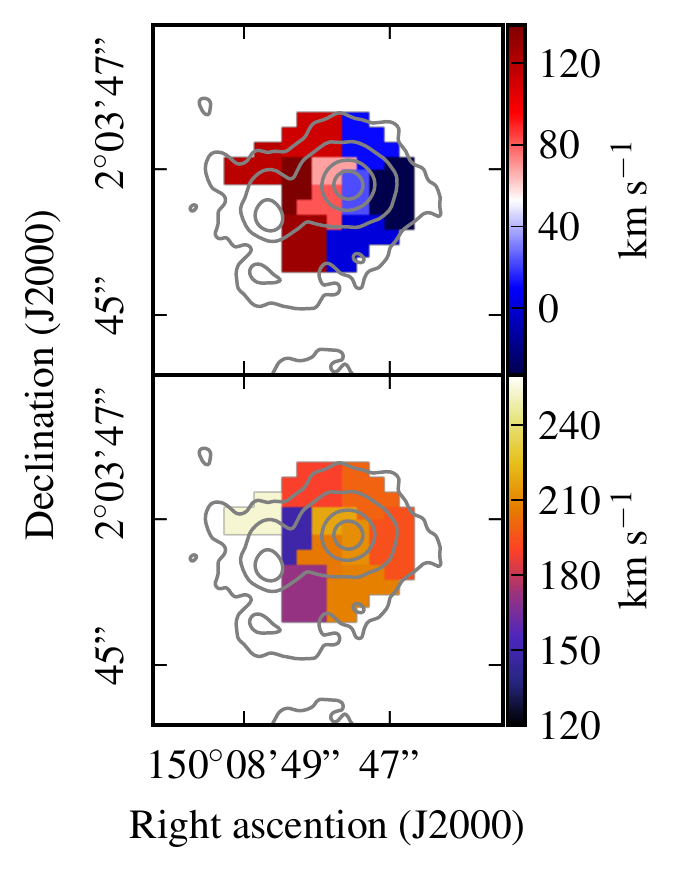}
 \includegraphics[width=4.45cm]{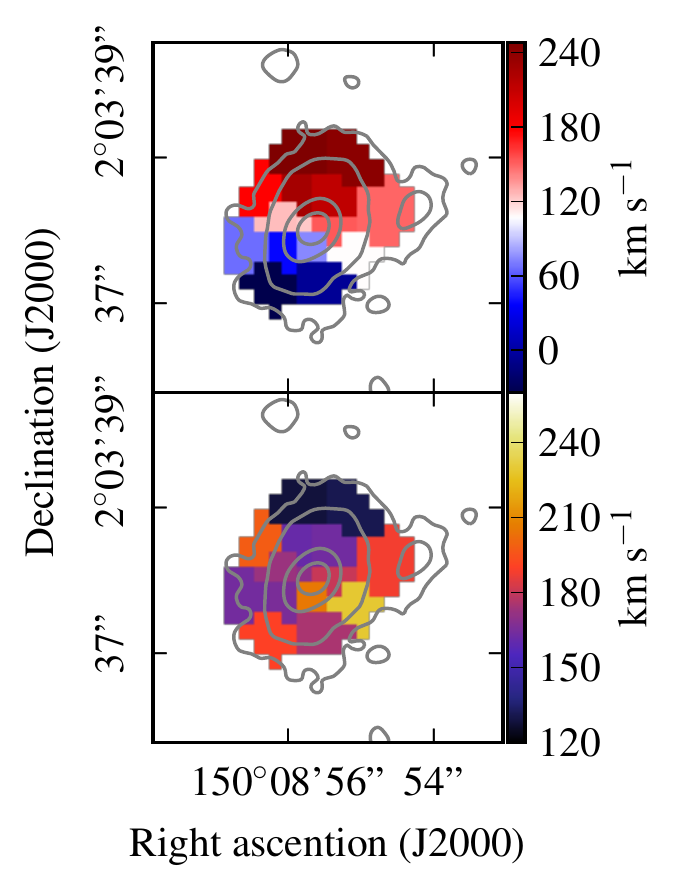}
 \caption{Stellar kinematics for CGr30-98 (left) and CGr30-71 (right). Velocity fields and velocity dispersion maps are shown in the top and bottom panels, respectively, with HST contours overlaid in grey. The velocity fields are shown using the same redshift as we used for the extraction of \oii\ dynamics.
 }
 \label{stellarkin_maps}
\end{figure}

In Figure \ref{stellarkin_maps} we present the stellar velocity fields and stellar velocity dispersion maps for CGr30-98 and CGr30-71. The velocity fields are shown using the same redshift as we used for the extraction of \oii\ dynamics in order to better compare the overall velocity with respect to the structure. The velocity field ranges are centred on the systemic velocity of each galaxy.
The stellar kinematics is compatible with the galaxy morphology, that is, the kinematics and morphological major axes are aligned for the two galaxies.

\section{Results}
\label{results}

Using the various measurements enabled by our MUSE data, we can constrain the physical properties of the ionised gas such as its mass, its density and temperature, and the photo-ionisation mechanisms at play in the extended structure and in galaxies.

\subsection{Gas mass and star formation rate estimates}

Two methods can be used to determine the mass of gas involved in the ionised structure. The first makes use of \mgii\ absorption in background galaxies,
whereas the other relies on the SFR estimated from emission line measurements using the Kennicutt-Schmidt law.

\subsubsection{Total mass of the diffuse gas from \mgii\ absorption}
\label{mass_mgii}

\begin{figure}
 \includegraphics[width=9cm]{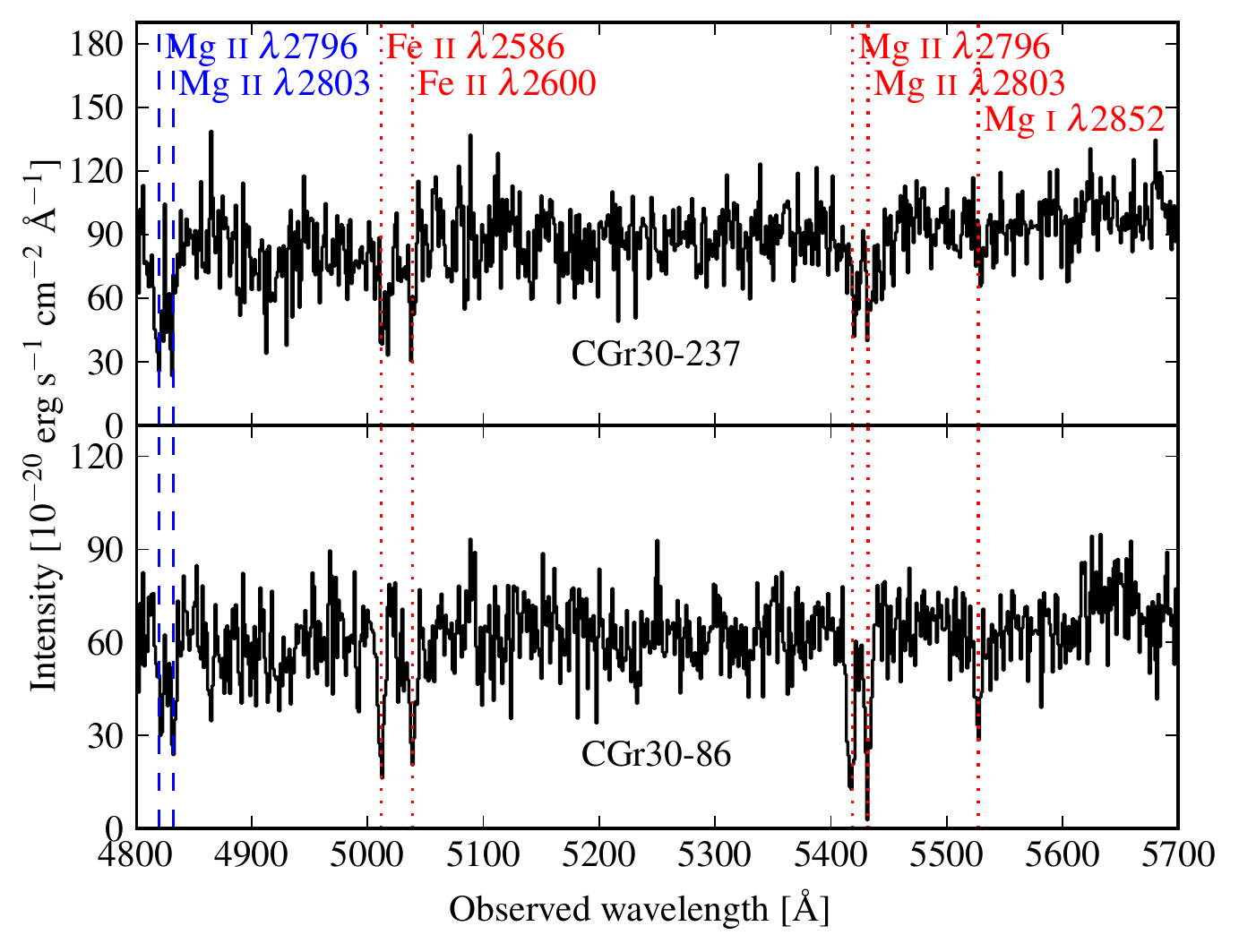}
 \caption{Spectra of the two background galaxies, CGr30-237 (top) and CGr30-86 (bottom), in the blue wavelength range of MUSE. The blue vertical dashed lines correspond to the absorption lines at the redshift of the group ($z\sim 0.724$), whereas the red vertical dotted lines correspond to the absorption lines at the redshift of the galaxies ($z\sim 0.938$).
 }
 \label{absorbers}
\end{figure}

Two background galaxies, CGr30-237 ($\text{RA}=150^{\circ}08'48.8''$, $\text{Dec}=2^{\circ}03'46.5''$) and CGr30-86 ($\text{RA}=150^{\circ}08'51.6''$, $\text{Dec}=2^{\circ}03'49.3''$) identified in Figure \ref{hst_members} with thick red circles, are located on the line of sight of the ionised structure and exhibit two absorption lines at $\lambda = 4834.66$ \AA\ and $\lambda = 4822.24$ \AA\ (cf. Figure \ref{absorbers}) corresponding to the \mgii\ doublet ($\lambda = 2796$ \AA\ and $\lambda = 2803$ \AA) at $z\sim 0.724$, the redshift of the group.
Given that these two galaxies are at $z = 0.938$, these lines are not likely to be associated with any self-absorbing material. The spectra do, however, show absorption at $z=0.938$ from the \mgii\ doublet, the \ion{Mg}{i} line ($\lambda = 2852$ \AA), and the \ion{Fe}{ii} doublet ($\lambda = 2586$ \AA\ and $\lambda = 2600$ \AA) associated with these two galaxies.

We used the relation between the \ion{H}{i} column density $N_{\ion{H}{i}}$ and the \mgii\ $\lambda 2796$ equivalent width at rest-frame $W_{\lambda 2796}$ of \citet{Menard:2009} to have a rough estimate of the extended gas column density on the line of sight of these two background galaxies,\begin{equation}
 N_{\ion{H}{i}} = 3.06\times 10^{19} \times W_{\lambda 2796}^{1.73}
 \label{column_density_mgii}
,\end{equation}
where $N_{\ion{H}{i}}$ is in atoms cm$^{-2}$ and $W_{\lambda 2796}$ is in Angstroms.

Spectra were extracted over circular apertures of four (CGr30-237) and three pixels (CGr30-86) radius. These spectra are shown in Figure \ref{absorbers}. They were adjusted around the \mgii\ lines using two Gaussian profiles and a flat continuum.
The \mgii\ $\lambda 2796$ equivalent width was computed for both galaxies.
For CGr30-237, we estimate $W_{\lambda 2796} = 3.21\pm 0.87$~\AA,\ which translates roughly into $N_{\ion{H}{i}} = (2.3 \pm 1.1) \times 10^{20}$~atoms~cm$^{-2}$. For CGr30-86, we estimate $W_{\lambda 2796} = 1.47\pm 0.77$~\AA, which translates roughly into $N_{\ion{H}{i}} = (6 \pm 5) \times 10^{19}$~atoms~cm$^{-2}$. 
The difference in column density between the two galaxies is compatible with the fact that CGr30-86 is located at the edge of the distribution, whereas CGr30-237 is more in the centre of the structure where \oii\ flux is stronger as well.

In order to infer a total gas mass for the structure, we assumed the column density of the diffuse gas to be constant over the whole structure. 
We used a covering factor equal to unity, which may represent an upper limit because the gas could be clumpy. However, since the background sources are extended, they may sample an area larger than potential gas sub-structures.
We used the column density of CGr30-237 since the detection is more robust and because it is better centred on the extended structure. We defined two components based on the kinematics of the whole extended structure (cf. section \ref{gaskin}). The size of the structure was inferred from the \oii\ flux map smoothed using a 0.28'' FWHM Gaussian extra filtering. When we include the 0.4'' filtering applied on cubes before map extraction, the resulting smoothing is around 0.5''. All pixels outside the main contour corresponding to the total \oii\ surface brightness detection limit at $3\sigma = 1.5\times 10^{-18}$~erg~s$^{-1}$~cm$^{-2}$~arcsec$^{-2}$ were discarded to obtain a cleaned map with a similar extent as the velocity field presented in Figure \ref{vfs}. Galaxies were not removed since the extended gas is observed beyond them.
When we count the number of pixels in each component and use a scale of 7.244 kpc/'' at $z=0.725$,
the total area is 5051 kpc$^2$ ($4.81 \times 10^{46}$~cm$^2$) and 2351 kpc$^2$ ($2.24 \times 10^{46}$~cm$^2$) for the northern and southern components, respectively.
The HI gas mass is therefore estimated to be $(9.2 \pm 2.8) \times 10^9$~M$_\odot$ for the northern component and $(4.3 \pm 1.3) \times 10^9$~M$_\odot$ for the southern component.
When we assume that the gas is composed of 75\% hydrogen and 25\% helium, the total gas mass is $(1.2 \pm 0.4) \times 10^{10}$~M$_\odot$ for the northern component and $(5.7 \pm 1.7) \times 10^9$~M$_\odot$ for the southern component.

The quoted uncertainties on these mass estimates are lower limits because they only reflect the uncertainties on the equivalent widths because the signal-to-noise ratio in the spectra is rather
low. One additional source of uncertainties is the determination of the surface, which strongly depends on the surface brightness threshold. A threshold of $2.0\times 10^{-18}$~erg~s$^{-1}$~cm$^{-2}$~arcsec$^{-2}$ ($3.0\times 10^{-18}$~erg~s$^{-1}$~cm$^{-2}$~arcsec$^{-2}$) instead of $1.5\times 10^{-18}$~erg~s$^{-1}$~cm$^{-2}$~arcsec$^{-2}$ would lead to surfaces of $\sim 4670$~kpc$^2$ ($\sim 3990$~kpc$^2$) and $\sim 1880$~kpc$^2$ ($\sim 1120$~kpc$^2$) for the northern and southern components, respectively, or in other words, to uncertainties of the order of 10 - 20 \%\ for the northern component and 20 - 50 \%\ for the southern component.
We also extrapolated the column density over the whole structure from one peculiar measurement. Using CGr30-86 instead of CGr30-237 would have led to four times lower masses. Finally, we used a canonical oversimplified relation to infer column densities from \mgii\ equivalent widths, and the dispersion around this relation is high. For these reasons, we estimate that the final uncertainty on these mass estimates is of the order of 50\%.

\subsubsection{Total mass of the diffuse gas from the star formation rate}
\label{mass_sfr}

\begin{figure}[ht]
\centering \includegraphics[width=8.3cm]{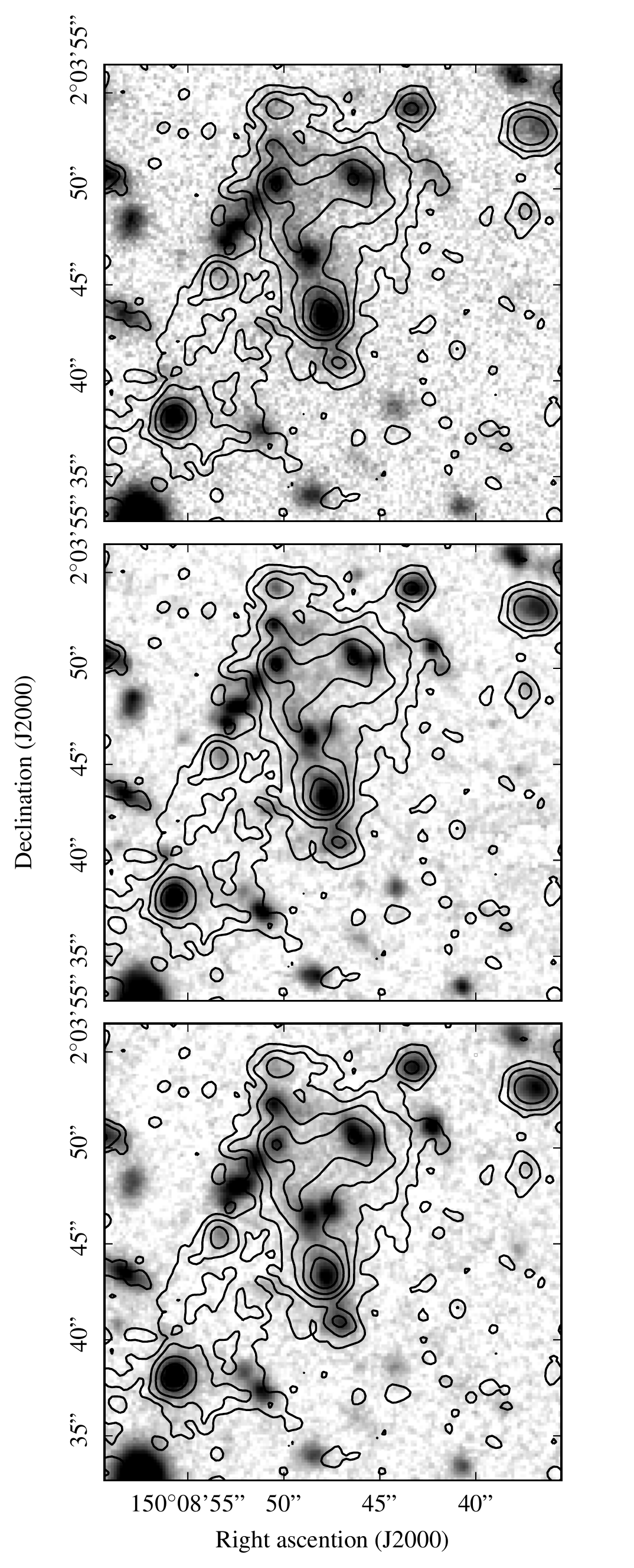}
 \caption{U-band MegaCam/CFHT image (top), B-band (middle), and V-band (bottom) SubprimeCam/Subaru images over the structure using a logarithmic scale and arbitrary units. The same logarithmic \oii\ contours as displayed in Figure \ref{hst_members} are overlaid on each image.}
 \label{broadband_ubv}
\end{figure}

We can infer an SFR from emission lines assuming they are related to ionisation by young stars and using the Kennicutt-Schmidt relation \citep{Kennicutt:1998b}, which provides a correlation to infer the gas surface density from the SFR surface density:
\begin{equation}
 \frac{\Sigma_{\text{SFR}}}{[\text{M}_\odot \text{ yr}^{-1} \text{ kpc}^{-2}]} = 2.5\times 10^{-4} \left( \frac{\Sigma_{\text{gas}}}{[\text{M}_\odot \text{ pc}^{-2}]} \right)^{1.4}
 \label{kennicutt_sfr_gas}
.\end{equation}
In our case, several mechanisms may be responsible for the ionisation of extended gas (see section \ref{ionisation_sources}). Since blue rest-frame broad-band images suggest the presence of young stars in the diffuse region (cf. Figure \ref{broadband_ubv}), we used this relation to have a second rough estimate of the gas mass in the diffuse structure.

In a first step, we used the \hb\ line to estimate the SFR. We assumed that the gas is homogeneously distributed over the region where \oii\ is detected, a null self-extinction, and temperature and density conditions so that \ha\ flux is 2.863 times \hb\ flux. The \citet{Kennicutt:1998} relation between SFR and \ha\ flux therefore leads to
\begin{equation}
 \frac{\text{SFR}_{Balmer}}{[\text{M}_\odot \text{ yr}^{-1}]} = 2.3 \times 10^{-41} \frac{\text{L}(\text{H}\beta)}{[\text{erg s}^{-1}]}
 \label{kennicutt_sfr_ha_hb}
.\end{equation}

The \hb\ flux was determined in an integrated spectrum computed for the northern component. This line is not detected in the southern component. To ensure that the line emission stems from the extended gas alone, the spectra were integrated over apertures defined on extended gas regions in section \ref{apertures}.
An average Galactic extinction of $E(B-V) = 0.0178$ in the direction of the structure was taken into account, as we described in section \ref{ionised_measurements}.
We neglected any in situ dust extinction as it is almost impossible to constrain from the Balmer decrement and is assumed to be much lower in the extended diffuse gas than in galaxies (cf. section \ref{balmer_extinction}).
The luminosity was computed using the luminosity distance ($l_d= 4.4\times 10^{3}$~Mpc) at $z\sim 0.725$. The SFR was derived using equation \ref{kennicutt_sfr_ha_hb}, the average SFR surface density was then computed by dividing the SFR by the area of the apertures used, and finally, the gas surface density was derived from equation \ref{kennicutt_sfr_gas}.
The total gas mass was estimated by taking the whole area into account, that is, including the area with galaxies where we can expect some extended gas to be present, as we did for the \mgii\ absorption line:
\begin{equation}
 \text{M}_{\text{gas}}= \Sigma_{\text{gas}} \times Area
 \label{gas_mass}
.\end{equation}
We found an SFR of $12\pm 2$~M$_\odot$~yr$^{-1}$ and a total gas mass of $(2.5 \pm 0.3) \times 10^{10}$~M$_\odot$ for the northern component. The uncertainties were derived from uncertainties on integrated fluxes.

Because \hb\ is not detected over the whole structure, we also used the \oii\ flux to infer the total gas mass, even if its flux depends on metallicity and on the ionisation parameter in addition to the number of ionising photons.
We were able to determine \oii\ fluxes for each component separately. As for \hb,\, we used spectra integrated over apertures defined on extended gas regions in section \ref{apertures}, the same Galactic extinction, and a null in situ dust extinction to compute the average \oii\ surface brightness.
The link between the \oii\ flux and the SFR was made using the \citet{Kennicutt:1998} relation,\begin{equation}
 \frac{\text{SFR}_{[\ion{O}{ii}]}}{[\text{M}_\odot \text{ yr}^{-1}]} = (1.4\pm 0.4) \times 10^{-41} \frac{\text{L}([\ion{O}{ii}])}{[\text{erg s}^{-1}]}
 \label{kennicutt_sfr_oii}
,\end{equation}
from which we can deduce the SFR surface density. Equations \ref{kennicutt_sfr_gas} and \ref{gas_mass} were used to derive the gas surface density and total gas mass.

We find an SFR of $37 \pm 11$~M$_\odot$~yr$^{-1}$ and $5.0 \pm 1.5$~M$_\odot$~yr$^{-1}$ and gas masses of $(5.6 \pm 1.2) \times 10^{10}$~M$_\odot$ and $(1.1 \pm 0.3) \times 10^{10}$~M$_\odot$ for the northern and southern components, respectively. In this case, the uncertainties are dominated by the uncertainty on the slope of equation \ref{kennicutt_sfr_oii}.

We also attempted to derive the mass directly on a pixel-per-pixel basis from the \oii\ flux map. We determined the gas mass inside each pixel outside galaxies using the above recipes. We then corrected for the missing pixels, assuming that the average diffuse gas density is identical in and outside galaxies. We found gas masses of $(4.0 \pm 0.9) \times 10^{10}$~M$_\odot$ and $(1.2 \pm 0.3)\times 10^{10}$~M$_\odot$ for the northern and southern components, respectively, which is consistent with the above estimates within the uncertainties.

The gas mass inferred from \oii\ is higher than the one deduced from \hb\ by a factor $\sim 2$.
Although these masses are higher than those estimated from the \mgii\ equivalent widths, the order of magnitude is quite consistent.
The masses from the SFR may be overestimated for two main reasons. First, the apertures were defined on the brightest regions, and second, a fraction of the ionisation may not be due to star formation (see section \ref{ionisation_sources}). In addition, physical conditions (density and temperature) in the diffuse ionised gas may be different from the conditions expected inside galaxies, where the Kennicutt-Schmidt relation is commonly used (see section \ref{electron_density_temperature}).

During this analysis, we also evaluated the total \oiiab, \hb,\ and \oiiiab\ line fluxes. We extracted an integrated spectrum using all spaxels of the northern and southern components where \oii\ is above the 3$\sigma$ threshold, that is, those displayed in the velocity fields presented in Figure \ref{vfs}.
We found a total \oii\ doublet flux of $(11.1 \pm 0.2) \times 10^{-16}$~erg~s$^{-1}$~cm$^{-2}$, a total \oiiiab\ doublet flux of $(6.1 \pm 0.4) \times 10^{-16}$~erg~s$^{-1}$~cm$^{-2}$ , and a total \hb\ line flux of $(2.6 \pm 0.2) \times 10^{-16}$~erg~s$^{-1}$~cm$^{-2}$.

\subsection{Electron density and temperature}
\label{electron_density_temperature}

In principle, the electron density can be derived from the ratio of the two lines of the \oii\ doublet. Using a doublet such as \oii\ is very convenient since it is not sensitive to extinction.
In practice, when the doublet is not well resolved, measuring both lines independently is challenging. The spectral resolution of the MUSE data is sufficient to separate the two lines if the line width remains lower than the doublet separation: the LSF FWHM corresponds to $\delta v \sim 120 $~km~s$^{-1}$ at the wavelength of the doublet at $z=0.725,$ whereas the wavelength separation of the \oii\ doublet corresponds to a shift in velocity of 220~km~s$^{-1}$. However, such a broadening can be reached either through local mechanisms (turbulence) or by large-scale motion that is smeared out when integrating over apertures.
For most of the extracted spectra, the doublet is not correctly resolved. For galaxies, it is mainly due to large-scale motions, whereas for extended gas, it seems to be limited by the local turbulence.

However, spectra over seven apertures show a well-resolved doublet: galaxies CGr30-59, CGr30-69, CGr30-110, CGr30-119, CGr30-227, CGr30-228, and the diffuse region CGr30-F.
For the six galaxies, which include five low-mass star-forming ones ($\lesssim 10^{10}$~M$_\odot$), the  $\frac{[\ion{O}{ii}]\lambda 3729}{[\ion{O}{ii}]\lambda 3726}$ ratio is clearly above unity, equal to 1.4 on average, with a standard deviation of 0.1, which indicates rather low electron densities.
We used the n-levels atom calculations of \texttt{tstsdas/Temden} IRAF task to determine the corresponding electron density assuming a temperature $T_e = 10000$ K, which is typical of star-forming regions. We found an electron density of n$_e \sim 55$ electrons cm$^{-3}$, which is also typical of \ion{H}{ii} regions in the Local Universe.

The ratio is equal to 1.9 for CGr30-F, which is above the theoretical limit of 1.5 allowed for star-forming regions with typical temperatures. This may indicate that the electron density is very low and that the temperature in the diffuse gas may be significantly lower than in star-forming regions.

In principle, the electron temperature can be determined using the $\frac{\text{\oiiia} + \text{\oiiib}}{\text{\oiiic}}$ ratio. Using our data for the brightest \oiiiab\ emitters and a 3$\sigma$ threshold on the \oiiic\ line, we can only infer an upper limit on the electron temperature of T$_e <30000$ K using the same \texttt{tstsdas/Temden} IRAF task.

\subsection{Extinction from the Balmer decrement}
\label{balmer_extinction}

\begin{table*}[t]
\caption{
Physical parameters and diagnostic line ratios derived in extended gas regions and group galaxies.
Column 1: MUSE ID for the COSMOS-Group 30. Alphabetic IDs correspond to apertures on diffuse gas, numeric IDs correspond to galaxies. Column 2: Median velocity dispersion in~km~s$^{-1}$. Column 3: Extinction derived from the Balmer decrement. For five regions, line flux measurements lead to a negative extinction that remained compatible with a null extinction ($\geq 0$) given the uncertainties,
however. Column 4: Electron density in atoms cm$^{-3}$. Columns
5-7: The three line diagnostics R23, O32, and \oiii/\hb,\ respectively.
}
\begin{center}
 \begin{tabular}{ccccccc}
\hline
CGr30-ID & $\sigma_v$ [km s$^{-1}$] & $E(B-V)_{Balmer}$ [mag] & n$_e$ [cm$^{-3}$] & R23 & O32 & \oiii/\hb\\
 (1) & (2) & (3) & (4)  & (5) & (6)  & (7) \\
\hline
A        & $195 \pm 77$    &  -                        &  -       & $>2.0$                    & $<1.15$                   & -                         \\
B        & $243 \pm 35$    &  -                        &  -       & $>1.6$                    & $<4.44$                   & -                         \\
C        & $99 \pm 45$     &  -                        &  -       & $>2.5$                    & $<1.02$                   & -                         \\
D        & $99 \pm 13$     & $>0.41$                   &  -       & $5.5^{+1.9}_{-0.9}$       & $0.22^{+0.12}_{-0.12}$    & $1.02^{+0.69}_{-0.53}$    \\
E        & $145 \pm 42$    & $\geq 0$                  &  -       & $5.1^{+0.5}_{-0.4}$       & $<0.21$                   & $<1.01$                   \\
F        & $103 \pm 18$    & $\geq 0$                  & $<20$    & $4.2^{+0.4}_{-0.3}$       & $0.30^{+0.04}_{-0.04}$    & $0.97^{+0.16}_{-0.13}$    \\
G        & $84 \pm 66$     &  -                        &  -       & $>4.2$                    & $0.81^{+0.21}_{-0.18}$    & $>1.57$                   \\
H        & $91 \pm 78$     &  -                        &  -       & $>2.0$                    & $<1.26$                   & -                         \\
I        & $75 \pm 57$     &  -                        &  -       & $>1.1$                    & $<1.17$                   & -                         \\
\hline
59       & $82 \pm 13$     & $0.32^{+0.12}_{-0.11}$    & 44       & $4.2^{+0.1}_{-0.1}$       & $0.56^{+0.02}_{-0.02}$    & $1.52^{+0.05}_{-0.05}$    \\
61       & $94 \pm 98$     &  -                        &  -       & $>0.3$                    & $<261.06$                 & -                         \\
68       & $162 \pm 92$    &  -                        &  -       & $>2.0$                    & $<1.26$                   & -                         \\
69       & $58 \pm 10$     &  -                        & 145      & $>5.2$                    & $0.76^{+0.15}_{-0.14}$    & $>1.97$                   \\
71       & $144 \pm 18$    & $0.33^{+0.20}_{-0.18}$    &  -       & $5.8^{+0.3}_{-0.3}$       & $3.19^{+0.22}_{-0.17}$    & $4.45^{+0.23}_{-0.19}$    \\
82       & $158 \pm 40$    & $\geq 0$                  &  -       & $3.7^{+0.6}_{-0.4}$       & $<0.78$                   & $<2.05$                   \\
98       & $158 \pm 38$    &  -                        &  -       & $>2.7$                    & $<0.76$                   & -                         \\
105      & $83 \pm 23$     & $\geq 0$                  &  -       & $1.5^{+0.1}_{-0.1}$       & $0.29^{+0.05}_{-0.05}$    & $0.33^{+0.06}_{-0.05}$    \\
110      & $64 \pm 50$     & $0.09^{+0.13}_{-0.12}$    & 42       & $1.0^{+0.1}_{-0.1}$       & $<0.22$                   & $<0.18$                   \\
119      & $80 \pm 34$     & $0.49^{+0.40}_{-0.34}$    & 106      & $2.3^{+0.3}_{-0.2}$       & $0.33^{+0.08}_{-0.08}$    & $0.53^{+0.16}_{-0.13}$    \\
131      & $220 \pm 63$    & $\geq 0$                  &  -       & $4.6^{+0.4}_{-0.4}$       & $0.27^{+0.05}_{-0.04}$    & $0.96^{+0.14}_{-0.13}$    \\
227      & $64 \pm 4$      & $0.18^{+0.17}_{-0.16}$    & $<20$    & $4.1^{+0.2}_{-0.2}$       & $0.63^{+0.03}_{-0.03}$    & $1.59^{+0.10}_{-0.09}$    \\
228      & $67 \pm 14$     & $>1.24$                   & 23       & $3.7^{+0.6}_{-0.4}$       & $0.55^{+0.06}_{-0.06}$    & $1.32^{+0.26}_{-0.17}$    \\
\hline
 \end{tabular}
\end{center}
\label{ext_sfr_mass_table}
\end{table*}

Extinction in the galaxy group can be estimated from the measurement of at least two Balmer lines. In most of galaxy spectra, the \hb\ line is clearly identified. In some cases, \hg\ could also be measured, and in a few cases, the Balmer lines were even detected down to H10 (see Table \ref{lines_table}).
We decided to use only \hb\ and \hg\ to perform a uniform analysis on most of the sources and because high-energy transitions are more affected by noise and therefore would not lead to better extinction estimates. We thus derived the colour excess from the ratio of \hg\ over \hb\ for each region showing both lines following the recipe from \citet{Momcheva:2013}:
\begin{equation}
 E(B-V)_{Balmer} = \frac{-2.5}{k(\text{H}\beta) - k(\text{H}\gamma)} \times \log\left( \frac{(\text{H}\gamma/\text{H}\beta)_{int}}{(\text{H}\gamma/\text{H}\beta)_{obs}} \right)
\label{colour_excess}
,\end{equation}
where we used the \citet{Calzetti:2000} extinction curve to estimate $k(\text{H}\beta)=4.60$ and $k(\text{H}\gamma)=5.12$ and assumed an intrinsic Case B Balmer recombination ratio of $(\text{H}\gamma/\text{H}\beta)_{int} = 0.468$. These values are appropriate for \ion{H}{ii} regions of temperature $T_e = 10000$ K and electron densities n$_e = 100$~cm$^{-3}$ \citep{Osterbrock:1989, Dopita:2003b}, which are close to the estimates we obtained in section \ref{electron_density_temperature}.

We used this recipe to estimate the reddening for both galaxies and regions in the extended structure. A Monte Carlo approach was used to propagate the uncertainties on the Balmer line fluxes quoted in Table \ref{lines_table}, assuming normal distributions.
When no Balmer line was detected, it was not possible to determine the extinction. However, when \hb\ was the only Balmer line, a lower limit was computed using the 3$\sigma$ detection threshold on the \hg\ line and the 1$\sigma$ uncertainty on \hb. These values compare within the 1$\sigma$ uncertainties quoted in Table \ref{ext_sfr_mass_table} to the values derived from SPS fitting (see section \ref{global}) for galaxies, except for the three galaxies where Balmer estimates are negative but compatible with a null extinction within the 1$\sigma$ uncertainties.
Balmer extinctions were measured for three apertures only in the extended ionised gas region.
One of them (CGr30-F) is only weakly constrained. The two others would be negative, but are compatible with a null extinction within the 2$\sigma$ (CGr30-F) and 3$\sigma$ (CGr30-E) uncertainties. This suggests that temperature and density conditions in the extended medium may differ from those in galaxies.
We therefore decided in the following analysis to not correct for extinction in the extended gas or in galaxies, even if this is probably better justified in the extended gas since there may be less dust. However, we stress that the highest value of extinction measured in galaxies, from SED fitting or Balmer estimates, is equal to only $E(B-V)=0.62$. This leads to a factor of two between \oiiia\ and \oiiab\ for the reddest and bluest lines,
respectively. This choice has therefore a low impact on line ratios and avoids adding any trend that may bias the interpretation.

\subsection{Line diagnostics}
\label{diagnostics}

Several line ratios are used as proxies to determine the oxygen abundance of ionised gas: [\ion{N}{ii}]/H$\alpha$, [\ion{S}{ii}]/H$\alpha$, \oiii/\hb, R23, p3, O32, \oiiic/\oiiia, \neiii/\oii,\, etc. \citep[see e.g.][]{Melbourne:2002, Bianco:2016}. Most of these metallicity indicators suffer some degeneracy, mainly with the degree of ionisation, which is usually solved by combining several of them, for example, with the BPT diagram \citep{Baldwin:1981}.

From the MUSE data, we were able to derive R23, \oiii/\hb, O32, \oiiic/\oiiia,\ and \neiii/\oii. However, since the \neiii\ and \oiiic\ lines are intrinsically faint and marginally detected in our spectra (see Table \ref{lines_table}), we focused on R23 and O32, for which constraints can be provided for most of the galaxies and regions since \oii\ is strong. We also investigated \oiii/\hb,\ which can be constrained for a few galaxies and regions.

The R23 parameter, which is defined as
\begin{equation}
 \text{R}23 = \frac{[\ion{O}{iii}]\lambda 5007 + [\ion{O}{iii}]\lambda 4959 + [\ion{O}{ii}]\lambda 3729 + [\ion{O}{ii}]\lambda 3726}{\text{H}\beta}
 \label{r23_eq}
,\end{equation}
is a good metallicity indicator for star-forming regions where stellar photo-ionisation is the main ionisation source. However, it is degenerate with respect to the oxygen abundance \citep[e.g.][]{McGaugh:1991, Nagao:2006, Kewley:2008} and peaks at $\sim 10$ when $12+log(O/H)\sim 8$.
At fixed oxygen abundance, the ionisation parameter $q$, defined as the ratio between the flux of the ionising photons and the number density of hydrogen atoms, can be inferred from the ratio of the \oiii\ lines over the \oii\ doublet:
\begin{equation}
 \text{O}32 = \frac{[\ion{O}{iii}]\lambda 5007 + [\ion{O}{iii}]\lambda 4959}{[\ion{O}{ii}]\lambda 3729 + [\ion{O}{ii}]\lambda 3726}
.\end{equation}
Therefore, using these two proxies enables us to constrain both the ionisation degree and the gas-phase metallicity. When shocks or AGN contribute significantly to the ionisation, these two proxies can still provide useful constraints.
Last, the \oiii/\hb\ proxy, defined as
\begin{equation}
 \text{\oiii/\hb} = \frac{\text{\oiiia} + \text{\oiiib}}{\text{\hb}}
,\end{equation}
is commonly used in BPT diagrams to distinguish between stellar, shock, and AGN photo-ionisation regimes.
In our study, the \oiii\ and \hb\ line fluxes have similar levels and are lower than the \oii\ doublet flux, which is typical for \ion{H}{ii} regions. However, when photo-ionisation is due to an AGN, the \oiii\ flux may become higher than that of \hb, which may lead to an increase of both \oiii/\hb\ and R23 proxies.

For the three line diagnostics, we used $\text{\oiiia} + \text{\oiiib}$ rather than $\text{\oiiia}$ in order to increase the signal-to-noise ratio of the measurements, since the relative flux of these two lines is constant.
The three line diagnostics were derived for all the apertures (see Table \ref{ext_sfr_mass_table}).
The uncertainties on each line flux were propagated through a Monte Carlo approach assuming normal distributions.
When the \hb\ (\oiii) line was not detected, a lower (upper) limit on R23 (O32) was inferred from the 3$\sigma$ detection threshold in each spectrum and the uncertainties associated with
the measurable line fluxes.
Similarly, when \oiii\ (\hb) was detected alone, a lower (upper) threshold on \oiii/\hb\ was estimated, and no value was computed when none of these lines was detected.

In Figure \ref{O32_R23} we show O32 as a function of R23 for all the apertures defined in section \ref{apertures} for both extended ionised gas regions (red dots) and group galaxies (blue dots).
On the one hand, it is clear that these proxies are hardly constrained for most of the extended gas regions. Only CGr30-F and CGr30-D extended gas regions have tight constraints on both O32 and R23 (see also Table \ref{ext_sfr_mass_table}). For the CGr30-G region, the O32 proxy can also be measured, and the large lower limit on R23 provides a good constraint for the typical range of R23. The CGr30-E region has R23 constrained and a low upper limit on O32, which also provide good constraints in the diagram.
Other extended gas regions only have lower limits on R23 and upper limits on O32 because of poor \hb\ and \oiii\ measurements,
which induce a poor constraint on their position in this diagram.
On the other hand, the position of galaxies is relatively well constrained in this diagram, except for CGr30-61, CGr30-68 (which is exactly at the same location as one extended gas region), and CGr30-98, for which neither R23 nor O32 could be measured. Whereas the CGr30-61 spectrum cannot provide any useful constraint, for CGr30-68 and CGr30-98, both parameters still remain better constrained than most of the extended gas regions. For CGr30-69, R23 was not measured, but the associated upper limit is quite high, and for CGr30-110 and CGr30-82, O32 was not measured, but the lower limits provide some constraints.

The \oiii/\hb\ was computed for ten galaxies and four apertures in the diffuse gas structure. For galaxies it is lower than two, except for CGr30-71, for which it reaches a value of 4.45, and for CGr30-69, which has a lower limit only because the spectrum has a low signal-to-noise ratio. For extended gas regions, it is lower than one, except for region CGr30-G, which has a lower limit of $\sim 1.6$.

\section{Interpretation}
\label{interpretation}

\subsection{Sources of photo-ionisation}
\label{ionisation_sources}

The photo-ionisation of the extended structure can have diverse origins. In typical \ion{H}{ii} regions, photo-ionisation is mainly due to young stars. In the region covered by diffuse gas, there might be some diffuse light in U-, B-, and V-band images (Figure \ref{broadband_ubv}). These bands correspond to rest-frame wavelengths bluer than \oii\ at the redshift of the galaxy group. This diffuse light may indicate the presence of young stars. However, owing to the low significance of the detection, these images do not rule out a scenario where photo-ionisation is driven by other mechanisms that most likely act during gravitational interactions. Galaxy interactions may induce shocks through the gas flows that are caused by the merger process \citep{Rich:2015}. In dense environments, shocks can be further induced by the collision between gas clouds in a galaxy and in the intra-group medium, as is the case for Stephan's Quintet \citep{Rodriguez-Baras:2014}. In this compact group, the intra-group gas is some interstellar medium that is tidally stripped during previous interactions. Photo-ionisation can be powered by AGN as well, which can be fuelled with gas driven inwards by tidal forces during mergers.
Unusual excitation mechanisms by fast particles \citep[e.g.][]{Ferland:2009} can also induce bright \oii\ with respect to \oiii\ in cool-core groups and clusters.
In order to distinguish between these ionisation sources, we used the line diagnostics defined in section \ref{diagnostics}.

\begin{figure*}
\centering \includegraphics[width=18cm]{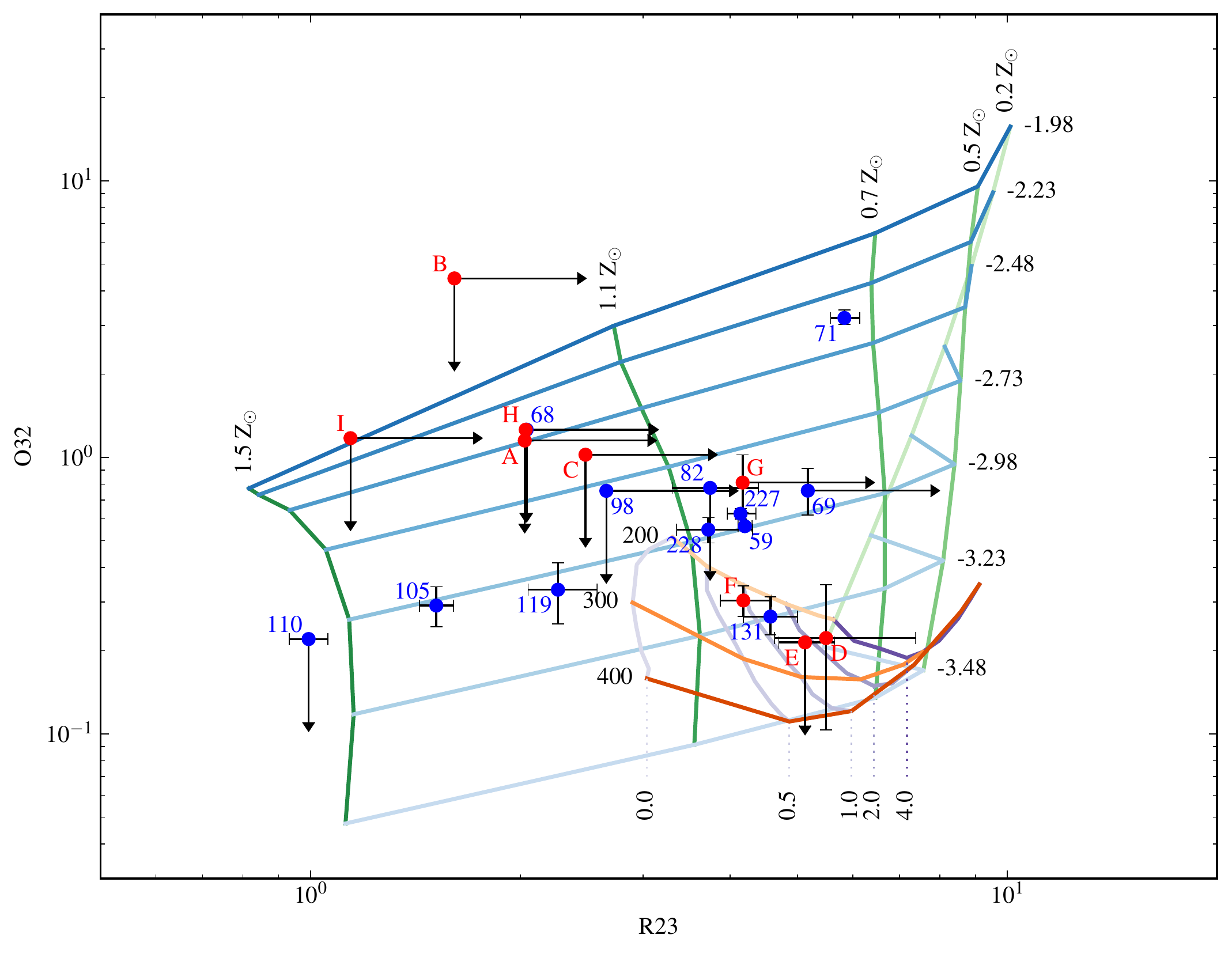}
\caption{O32 vs. R23 line ratio diagnostic diagram. Blue and red dots correspond to galaxies and extended gas regions, respectively. Their IDs are also indicated. Photo-ionisation models from MAPPINGS V \citep{Sutherland:inprep} are shown with blue (abundance) and green (ionisation parameter) lines. Shock models from MAPPINGS III \citep{Allen:2008} obtained without precursor assuming an LMC metallicity and a pre-shock density $n=1$~cm$^{-3}$ are shown with purple (magnetic parameter in $\mu$G) and orange (shock velocity in~km~s$^{-1}$) lines.}
 \label{O32_R23}
\end{figure*}

\subsubsection{Stellar radiation}
\label{stellar_radiation}

In order to constrain both the oxygen abundance and the ionisation parameter when the ionisation is due to stellar radiation, we have overplotted the grids from the photo-ionisation model MAPPINGS V \citep{Sutherland:inprep} on the R23 versus O32 diagram of Figure \ref{O32_R23}. 
We stress, however, that the MAPPINGS models may be less well
adapted for extended gas regions since they are designed for radiation-bounded \ion{H}{ii} regions.
We have used the \textsc{pyqz} 0.8.0 software \citep{Dopita:2013, Vogt:inprep} to extract the abundance versus ionisation grid with a typical pressure of \ion{H}{ii} regions ($\log{(P/k)}=5$) and with a Maxwell-Boltzmann distribution of electron energies. In Figure \ref{O32_R23} we provide the gas-phase abundance with respect to a solar oxygen abundance $\text{Z}_\odot= 8.69$ \citep{Asplund:2009} and the dimensionless ionisation parameter defined as $\text{U} = \text{q} / \text{c,}$ where c is the speed of light.

For galaxies, the ionisation parameter is typically below $\log{\text{U}}\sim -2.7,$ except for two galaxies (one with only an upper limit, and one with good constraints). For those with good constraints, the ionisation parameter seems rather constant around a value of $\log{\text{U}}\sim -3$, except for CGr30-71, for which $\log{\text{U}}\sim -2.5,$ and for CGr30-131, for which the inferred ionisation parameter is $\log{\text{U}}\sim -3.2$.
The gas-phase oxygen abundance spans a wide range for galaxies, from $\sim 1.5 \text{ Z}_\odot$ to $\sim 0.7 \text{ Z}_\odot$, with most of the galaxies around solar abundance.

For extended gas regions with no stellar counterpart, the ionisation parameter and oxygen abundances span almost the whole range that
is explored by MAPPINGS V simulations.
However, one of the four best-constrained regions (CGr30-G) is compatible with typical solar abundance and $\log{\text{U}}\sim -3$ observed in galaxies.
The three others are compatible with a low-ionisation parameter $\log{\text{U}}< -3.2$ and sub-solar oxygen abundance $0.7 \text{ Z}_\odot < \text{Z} < 1 \text{ Z}_\odot$, quite similar to galaxy CGr30-131. However, at the location of these three regions in the diagram, there is some degeneracy, and the abundance could also be lower than $0.2 \text{ Z}_\odot$.
If these regions are ionised by stellar radiation, this implies that there are only few ionising stars and a relatively high hydrogen density.
Therefore, extended gas could originate from galaxies themselves even if the gas seems less enriched than in galaxies. 

\subsubsection{Shocks}
\label{shocks}

The facts that extended gas regions are bright in \oii\ with respect to \hb\ and \oiii and that their typical velocity dispersion is higher than in galaxies suggest that a fraction of the ionisation might be due to shocks.
In order to test this hypothesis, we used the fast radiative shock model of \citet{Allen:2008}, which is based on MAPPINGS III. We used the SHOCKPLOT widget to study the impact of shocks on R23 and O32.
We tested various parameters and found that a relatively high R23 and a relatively low O32 can be obtained in shocks without precursors with an oxygen abundance from solar to $1/4$ solar, as observed in the Large Magellanic Cloud (LMC), and a pre-shock particle number density from 0.1 to 10 cm$^{-3}$.

In Figure \ref{O32_R23} we display the grid obtained for an LMC abundance and a pre-shock particle number density of 1 cm$^{-3}$ without precursor for various shock velocities and magnetic field parameters. Three regions (CGr30-F, CGr30-D, and CGr30-E) of the four extended gas regions with tight constraints in this diagram and one galaxy (CGr30-131) appear to match the location expected for such a shock, with a relatively low magnetic field and shock velocities of $\sim 200$ to 300~km~s$^{-1}$. These velocities are slightly higher than the velocity dispersion observed in extended gas regions, but they match that of CGr30-131 quite
well (see Table \ref{lines_table}), in agreement with expectations \citep{Rich:2011, Rich:2015, Rodriguez-Baras:2014}. Changing the abundance to solar only modifies the value of the magnetic field.
These observations therefore suggest that the extended gas in these regions could be ionised by shocks.
In this case, the abundance indicates that the extended gas could have a galactic origin.
The only galaxy for which the position in the R23 versus O32 diagram is compatible with a shock ionisation of the gas is CGr30-131. The high velocity dispersion observed in this galaxy, and most specifically, on the south-west side, supports this shock hypothesis.
This galaxy is embedded in the extended structure, and its orientation suggests that it might rotate almost orthogonally to the extended gas, which in this specific area leads to shocks between intra-group gas and gas inside the galaxy. At this specific location, the diffuse gas velocity dispersion is higher than 150~km~s$^{-1}$, which is compatible with this hypothesis.

If the gas is ionised by shocks, the \hb\ luminosity emitted per unit of surface area $L_{\text{\hb}}$ might scale with shock velocity $V_s$ and pre-shock number density $n$ \citep{Dopita:1996}:
\begin{equation}
 L_{\text{\hb}}=7.44 \times 10^{-6} \left( \frac{V_s}{100 \text{ km s}^{-1}}\right)^{2.41} \times \left( \frac{n}{\text{cm}^{-3}}\right)\text{ erg cm}^{-2} \text{ s}^{-1}
 \label{hbflux_shock}
.\end{equation}

For the northern component, the \hb\ luminosity is $5.2\times 10^{41}$ erg s$^{-1}$ over an area of $\sim 5000$~kpc$^2$ (see section \ref{mass_sfr}), hence the luminosity per surface unit is $1.1 \times 10^{-5}$ erg cm$^{-2}$ s$^{-1}$.
Assuming a typical shock velocity of $\sim 150$~km~s$^{-1}$, estimated from the mean diffuse gas velocity dispersion in the northern component, we infer a pre-shock number density of $\sim 1.1$~cm$^{-3}$, which is consistent with the constraints on R23 versus O32 from the models of \citet{Allen:2008}. 
The typical \ion{H}{i} column density inferred from \mgii\ absorption measurements is $2.3\times 10^{20}$ atoms cm$^{-2}$ (see section \ref{mass_mgii}). If the gas is distributed in a thin disc and mainly composed of neutral hydrogen, it would therefore have a thickness of $\sim 70$ pc. Slight changes of $\pm 20$~km~s$^{-1}$ in shock velocity induce variations in the thickness of around 50\%. The diffuse gas of the northern component is therefore consistent with being distributed in a relatively thin disc.

For the southern component, the line fluxes are not measured with enough accuracy to derive robust positions in the R23 versus O32 diagnostic diagram. However, it is worth noting that the three apertures defined on extended ionised gas all have lower velocity dispersions ($\lesssim 90$~km~s$^{-1}$) than those defined in the northern component ($\gtrsim 100$~km~s$^{-1}$), which may indicate that shocks are less relevant in the southern region.

The shock hypothesis cannot explain the presence of ionisation over a large area well, as is revealed by the three extended gas regions with line ratios that are compatible with shocks that are spread over the northern component.
If the shock source is limited in both time and space, ionisation would vanish too quickly when the shock front travels. Shock ionisation may last shorter than $10^6$~yr after the passage of a shock front \citep{Allen:2008}, whereas the typical travel time of a galaxy in the structure might be around $\sim 10^9$~yr. This means that the shock source has to be still active. One explanation is that the extended gas is continuously shocked by various galaxies that are currently located inside the structure.
Another caveat is that emission before the shock front is expected. At the redshift of COSMOS-Gr30, the shock and its precursor should be unresolved, but models without a precursor match our data
better.
The whole structure is probably not fully ionised by shocks. In addition, the line ratios available with our MUSE data remain consistent with photo-ionisation for all regions.

\subsubsection{AGN}
\label{agn_ionisation}

The spectrum of CGr30-71 (cf. Figure \ref{spectra_ex}, top) suggests that this galaxy hosts an AGN.
It displays broad emission lines in the nucleus (cf. Table \ref{ext_sfr_mass_table}), a bright \oiii\ line with respect to \oii\ and \hb,\ and shows \nev\ and \neiii\ emission lines.
The presence of an AGN is further confirmed by the detection of this galaxy in Chandra, IRAC, MIPS, and VLA data.
This galaxy is offset with respect to other galaxies in Figure \ref{O32_R23} because of its high O32 parameter, but this high value is mainly due to its strong \oiii\ emission. As described in section \ref{diagnostics}, it has a high \oiii/\hb\ line ratio, higher than 4, which is suggestive of an AGN for relatively massive galaxies \citep[e.g.][]{Kewley:2006, Rich:2015}.

\begin{figure}
 \includegraphics[width=9cm]{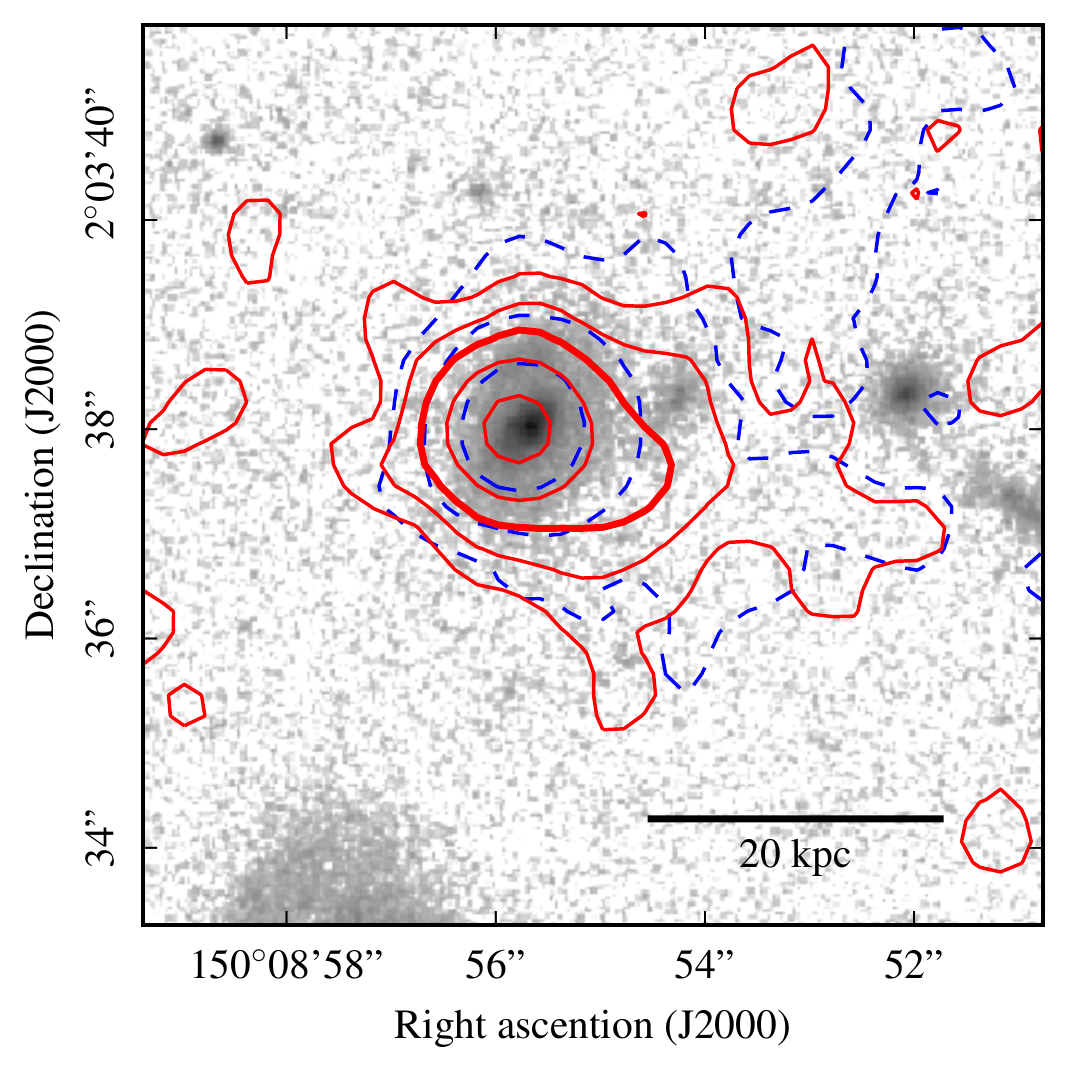}
 \caption{HST-ACS image (F814W filter, logarithmic scale, arbitrary unit) around the AGN-host galaxy CGr30-71. \oiiiab\ and \oiiab\ surface brightness contours, smoothed using a 0.2'' FWHM Gaussian, are displayed with red solid and blue dashed contours, respectively. The levels are identical for each doublet and correspond to 3.0, 7.6, 19.4, 49.5, and $125.9\times 10^{-18}$~erg~s$^{-1}$~cm$^{-2}$~arcsec$^{-2}$. \oiii\ is clearly stronger than \oii. The inner outflow-like feature observed in the \oiii\ flux map is highlighted by the thick red contour. The physical scale at the redshift of the structure is indicated at the bottom right.
 }
 \label{hst_oiii_oii}
\end{figure}

While the location of CGr30-G in Figure \ref{O32_R23} seems to be identical to that of normal galaxies, this region is the only extended region with \oiii/\hb\ higher than 1.5. This suggests
that it could be ionised by the AGN radiation rather than by stellar radiation. In addition, this region seems elongated along the minor axis of the AGN-host galaxy, which could be a signature of an anisotropic outflow that is produced by the AGN. This hypothesis is further supported by the detection in \oiii\ of an elongated structure that starts from the AGN and extends up to $\sim 2$'' ($\sim 15$~kpc) in the direction of CGr30-G (cf. Figures \ref{oiii_hb_oii_rgb} and \ref{hst_oiii_oii}). The end of this structure seems even more \oiii\ dominated than the AGN itself. The lower ionisation parameter of CGr30-G compared to that of the AGN would therefore be due to the larger distance with respect to the AGN \citep[see e.g.][]{Davies:2014}.

\subsubsection{Excitation by fast particles}
\label{cooling}

The ionised structure is observed in the densest part of a large-scale structure, which suggests that mechanisms similar to those responsible for the ionisation in cluster or group cores could occur.
Low \oiii/\oii\ ratios, such as those observed in some of the extended ionised regions of the northern component, are typical in cool-core groups and clusters that typically show weak \oiii\ compared to the other oxygen lines.
In local groups and clusters with cool cores, the ionised nebula shows filamentary structures that can extend over several tens of kpc \citep[e.g.][]{Conselice:2001, Hamer:2016}.
The filaments are believed to contain cold gas \citep[e.g.][]{Salome:2011} that is cooled from the intra-group medium and excited by fast-ionising particles \citep{Ferland:2009}, either from the surrounding hot gas \citep{Fabian:2011} or from the reconnection of gas phases within the filaments \citep{Churazov:2013}.

At the redshift of the observed structure, [\ion{O}{i}], \ha\ or [\ion{N}{ii}] emission lines are not observable in our MUSE data to fully compare its ionisation to models of ionisation by fast particles. In addition, radiative cooling may be correlated with X-ray emission of the intra-group medium. The available X-ray data (XMM and Chandra) do not show evidence of emission from an intra-group medium surrounding the galaxies or coincident with the ionised gas, which contradicts this hypothesis. Nonetheless, these data cannot confidently rule out the possibility of a weak intra-group medium.

\subsection{Global interpretation of the structure}

A variety of mechanisms responsible for the large-scale ionisation in this extended structure seem to
co-exist, as shown in section \ref{ionisation_sources}. These various mechanisms are probably triggered by some interaction in this over-dense region that may be located inside a cosmic web filament \citep{Iovino:2016}. Such a mixture (AGN, shocks, and star formation) has been observed previously, but on much smaller scales, in Stephan's Quintet \citep{Iglesias-Paramo:2012, Rodriguez-Baras:2014} or in U/LIRGs \citep[e.g.][]{Rich:2015},
for example, which are also driven by an interaction.
Based on our MUSE data, we can combine line ratio diagnostics measurements with large- and small-scale kinematics information to draw a coherent picture of the observed structure.
While the \oii\ flux map displays a continuous distribution, large-scale kinematics reveal two distinct components associated with the two most massive galaxies involved in the structure, namely CGr30-98 and CGr30-71 (see Figure \ref{vfs}). We interpret these two components separately in a first step.

\subsubsection{Northern component: large rotating disc}

The extended ionised gas in the northern component has a rather symmetrical velocity pattern with respect to CGr30-98 within a radius of 20~kpc.
On the one hand, the median line-of-sight velocity of CGr30-98 obtained from its stellar kinematics is compatible with an agreement better than 10~km~s$^{-1}$ to the ionised gas velocity at the same location (cf. Figures \ref{vfs} and \ref{stellarkin_maps}). The uncertainties associated with the ionised gas velocity at this position are relatively large ($\sim 20$~km~s$^{-1}$) because
of the large velocity gradient. Therefore gas and stars may be dynamically linked.
On the other hand, the kinematics position angle (with respect to north) obtained from stars ($90^\circ \pm 10^\circ$) is compatible with the morphological position angle of the major axis ($88^\circ$), but is orthogonal to that of the extended gas ($-17^\circ \pm 3^\circ$) within $\sim 3$'' around that galaxy.
Finally, the galaxy is offset by about 0.4'' ($\sim 2.9$~kpc) with respect to the centre of symmetry of the extended kinematics pattern.
These findings  may indicate (i) that ionised gas is radially outflowing from or (ii) inflowing into the galaxy, (iii) that the angular momentum of the galaxy has recently been redistributed with respect to that of the extended ionised gas, or (iv) that the gas has been trapped by the galaxy potential after the stellar disc built up.

If the gas is outflowing, we would probably expect a higher gas abundance. In addition, the shock hypothesis (cf. section \ref{shocks}) favours a distribution of the extended gas in a thin disc, which is unlikely for the outflow hypothesis. In case of an inflow, gas is usually accreted in the galaxy disc plane with not purely radial motions \citep{Bouche:2013, Bouche:2016}, therefore the gas distribution would probably be more elongated towards the galaxy major axis. 
For the two latter explanations, the extended gas may be distributed in a disc and trace the same gravitational potential as stars in the galaxy. In order to check these hypotheses, we computed the dynamical mass independently for these two tracers.
Assuming the extended ionised gas is rotating in a disc with an inclination of $60^\circ$ around CGr30-98, we inferred a dynamical mass of $2.7 \pm 0.3 \times 10^{11}$~M$_\odot$ at 13 kpc. We used the velocity difference between the extrema of the velocity field ($\Delta V = 560\pm 20$~km~s$^{-1}$), which are separated by a distance of $26\pm 2$~kpc.
We roughly estimated the mass from stellar dynamics using equation \ref{cap06}, where we assumed the effective radius to be that of the disc ($r_e = 3.9$~kpc) and the dispersion to be the median dispersion $\sigma\sim 200$~km~s$^{-1}$,
\begin{equation}
 \left( M/L \right)_{vir} = \beta r_e \sigma^2/(L G)
 \label{cap06}
.\end{equation}
We assumed $\beta = 5$, as suggested by \citet{Cappellari:2006}, and obtained a virial mass for CGr30-98 of $\sim 1.8\times 10^{11}$~M$_\odot$.
Using the bulge radius ($\sim 0.9$~kpc), the mass would be $\sim 4\times 10^{10}$~M$_\odot$.
Both gas and stellar dynamical masses have the same order of magnitude and are higher than the galaxy stellar mass ($\sim 4.3\times 10^{10}$~M$_\odot$). This supports the hypothesis that the extended ionised gas is bound to a dark matter halo associated with CGr30-98 and in rotation around this galaxy.
The difference observed in the orientation between gaseous and stellar major axis could be due to interactions or merger events.
The fact that the gas shows an enriched medium argues against a primordial gas reservoir origin. The non-primordial abundance of the gas could be due either to a reservoir fed by outflows from the galaxy or to gas stripped from galaxies during interactions or merger events, and trapped by the gravitational potential of the structure.

The main deviation from rotation around CGr30-98 is observed close to CGr30-131, where velocities are lower by around 250~km~s$^{-1}$ on average and by around 400~km~s$^{-1}$ at maximum with respect to surrounding extended gas. This velocity difference can be explained by the proper motion of CGr30-131 in the structure, inducing a line-of-sight velocity around $-200$~km~s$^{-1}$ with respect to the centre of symmetry of the extended gas velocity pattern, and by its internal kinematics, which leads to a $\sim 400$~km~s$^{-1}$ velocity shear across the galaxy. Given the mass, size, and inclination of CGr30-131, we expect the velocity amplitude due to its internal kinematics to be at least $200$~km~s$^{-1}$, which is in good agreement with the observed kinematics. These proper motions could be partially responsible for the high dispersion observed on and around CGr30-131, as we discussed in section \ref{gaskin}.
We also note that the motions of small galaxies (CGr30-227, CGr30-228, CGr30-119, CGr30-59, and CGr30-68) are compatible with the large-scale velocity pattern of the extended ionised gas within small deviations. In addition, the abundances found in these low-mass galaxies and in the extended ionised gas are compatible (cf. sections \ref{stellar_radiation} and \ref{shocks}). This may indicate that they are orbiting around the main galaxy CGr30-98 or that they have been formed from extended gaseous material rotating around this main galaxy. One explanation would be that they are tidal dwarf galaxies \citep[e.g.][]{Duc:2000, Duc:2004} induced by recent interactions that may partly be responsible for the presence of non-primordial gas at large scale in a disc decoupled from the stellar disc of CGr30-98. Recent interactions may have triggered recent starbursts, which is supported by the young age of stellar populations (cf. section \ref{continuum_removal}) and by strong high-order Balmer absorptions \citep[e.g.][]{GonzalezDelgado:1999} observed in some galaxies in the northern component (CGr30-131, CGr30-119), in the southern component (CGr30-71), and in CGr30-105. 

We stress that the kinematics of ionised nebula in the cores of clusters and groups commonly suggest rotation around the most massive galaxy and decoupling from that of the stars \citep{Hamer:2016}. The \ion{H}{i} gas mass estimates are also comparable, within the large uncertainties, to those found in cool-core clusters and groups. 
Given these similarities and those underlined in section \ref{cooling} (low O32 ratio, filamentary structure), another possible interpretation for the ionisation of the northern component is that cold gas is cooled from the inter-group medium that is excited by fast particles. This could mean that we are observing the formation of a group or cluster core. However, the lack of X-ray detection makes this hypothesis less probable than shocks and stellar photo-ionisation.

\subsubsection{Southern component: AGN outflow and tidal tails}

The southern component appears to be linked to CGr30-71, which hosts an AGN. The main ionised gas extent of this component is distributed in the direction of the major axis of CGr30-71, and its velocity is compatible with the velocity of the receding side of this galaxy as well as with the small galaxy CGr30-82. This could indicate some large-scale accretion of gas on CGr30-71. However, it seems to be composed of two filaments that link CGr30-71 and CGr30-82, suggesting that the ionised gas is in tidal tails. One interesting kinematics feature of these two filaments is that a velocity gradient is observed in a direction orthogonal to the filament. The velocities are higher on the west side by around 200~km~s$^{-1}$. 

A third filament leaves the galaxy along its minor axis with a projected velocity of around $-200\pm 50$~km~s$^{-1}$ with respect to the central velocity measured from stellar kinematics to avoid biases due to the AGN.
When we assume that this feature is an outflow from the AGN and that it is ejected orthogonally to the disc plane, which has an inclination of 40$^\circ$, the outflow velocity is around 250~km~s$^{-1}$ even at a distance of $\sim 35$~kpc ($\sim 5$'') from the galaxy. The velocity along this feature seems quite constant, except in the inner part, where the outflow has a velocity difference with respect to the centre that is much lower, of the order of 50~km~s$^{-1}$. It is even compatible with a null velocity from \oiii\ kinematics. A lower limit on the AGN activity duration of $\sim 0.1$~Myr can be inferred assuming the outflow travelled at the speed of light. With an outflow velocity of 250~km~s$^{-1}$ and if the outflow is orthogonal to the disc, the AGN activity would have last $\sim 200$~Myr.

\subsubsection{Interactions between the two components}

This is the first detection of such a structure at this and indeed at any other redshift. It is therefore quite intriguing that it displays two sub-structures at the same redshift that seem to be kinematically decoupled, however, and may be ionised by different mechanisms (shocks vs. AGN). One explanation would be that these sub-structures are decoupled today, but were in close interaction in the past. Such an interaction between the most massive galaxies may have expelled some gas from galaxies through tidal forces.
This interpretation is consistent with the presence of young stars that is suggested by faint signal in the B-band image (see Figure \ref{broadband_ubv}). This interaction may also explain the filamentary structure between two low-mass star-forming galaxies (CGr30-227 and CGr30-59) suggested in the \oii\ flux map, and
it may explain the symmetry of CGr30-227 and CGr30-228 with respect to CGr30-119.
Strong interactions could also have triggered past (and present) AGN in various galaxies (e.g. CGr30-71, CGr30-98 and CGr30-131), which could have powered some past outflows. These outflows could have been trapped in the potential of CGr30-98, depending on the geometry at these early times. These two explanations would account for gas being present over a large area on the sky.

If our interpretation of the kinematics of the northern component is correct, the existence of such a structure implies that a disc can be re-built through interactions, not necessarily with pristine gas, that is, previously enriched gas. Such an event could then lead to a kinematically decoupled core as observed in nearby elliptical galaxies \citep[e.g.][]{Krajnovic:2015} if the galaxies involved in the interaction do not merge in the end.

\section{Summary and conclusions}
\label{conclusion}

We reported the serendipitous discovery of a $10^4$~kpc$^2$ diffuse ionised gas structure in an over-dense region at redshift $z\sim 0.725$ in which a dozen galaxies are involved. This over-dense region is itself part of the larger scale COSMOS-Wall \citep{Iovino:2016}. This is the first time that such a large ionised gas structure has been observed at any redshift. The entire structure is clearly detected in \oii. In order to understand the origin and ionisation
source of this diffuse gas, we have studied the gas kinematics and emission lines measurements in the visible (mainly \hb\ and \hg\ Balmer lines, \oiiiab,\ and \oiiab) derived from deep MUSE data.

The kinematics revealed two main sub-structures that are separated by a velocity offset that we referred to as the northern and southern components. These two sub-structures seem linked to the two most massive galaxies that are embedded in the ionised gas. The massive galaxy in the southern component hosts an AGN.

We have estimated the mass of diffuse gas with two techniques, one using \mgii\ absorption features in background galaxies to infer the \ion{H}{i} column density, the other using either the \oii\ or \hb\ surface brightness and the Kennicutt-Schmidt law.
These methods give compatible orders of magnitude and lead to masses of diffuse ionised gas between $(1.2 \pm 0.6)$ and $(5.6 \pm 1.2) \times 10^{10}$~M$_\odot$ for the northern component and between $(0.6 \pm 0.3)$ to $(1.2 \pm 0.3) \times 10^{10}$~M$_\odot$ for the southern component.

We have used emission lines in MUSE spectra in order to understand the sources of ionisation within the structure. We extracted spectra from apertures defined on both the galaxies and the diffuse ionised gas regions for which we measured the line fluxes after continuum subtraction. Balmer line measurements were not accurate enough to derive any strong constraint on the extinction in the extended structure. Nevertheless, we studied the R23 and O32 line diagnostics for both the galaxies and the extended gas regions. We compared the positions of galaxies and extended gas regions in this diagram with photo-ionisation and shock models.

Our study suggests that the ionisation source of the northern component is a mixture of photo-ionisation and shocks without a precursor between the gas from at least one disc-like galaxy and some extended gas. The data are consistent with a shock velocity of about 200~km~s$^{-1}$, a low pre-shock density ($\sim 1\,\mathrm{cm}^{-3}$), an LMC abundance, and a weak magnetic field.
The shock hypothesis in the northern component is supported by the relatively high velocity dispersion, which is much higher than in typical star-forming regions. From this hypothesis, the gas is likely distributed in a plane.
Interestingly, regardless of the ionisation source, the diffuse gas is not primordial, indicating that it may have been extracted from the galaxies during interactions or AGN-related episodes of strong outflows.

Another explanation would be that fast particles are exciting cold gas, as can be observed in local groups and cool-core clusters. However, the absence of X-ray from an inter-group medium disfavours this hypothesis.

For the southern component, the ionisation source is less well constrained. However, at least one feature seems to be ionised by the AGN, based on its high \oiii/\hb\ ratio. This outflow-like feature extends up to $\sim 35$~kpc orthogonally to the disc of the AGN-host starting from its nucleus.

The large-scale \oii\ kinematics of the northern component suggest that the gas rotates around a massive galaxy in a plane almost orthogonal to the disc of the galaxy. Nevertheless, analysing the stellar kinematics leads to a virial mass of $1.8\times 10^{11}$~M$_\odot$, similar to the mass inferred from the large-scale ionised gas kinematics of about $2.7\times 10^{11}$~M$_\odot$ at a radius of 13 kpc.

The northern and southern components of the ionised gas structure are probably related.
Some gas could have been extracted from a previous interaction between the main galaxies, either from tidal interactions or AGN outflows induced by the interaction. Low-mass star-forming galaxies embedded in filament-like \oii\ structures may be tidal dwarfs.
We could therefore be observing the rebuilding of a disc after a merger episode.

\begin{acknowledgements}
This work has been carried out through the support of the ANR FOGHAR (ANR-13-BS05-0010-02), the OCEVU Labex (ANR-11-LABX-0060), and the A*MIDEX project (ANR-11-IDEX-0001-02), which are funded by the ``Investissements d'avenir'' French government program managed by the ANR.
BE acknowledges financial support from the ``Programme National de Cosmologie et Galaxies'' (PNCG) funded by CNRS/INSU-IN2P3-INP, CEA and CNES, France.
JB acknowledges support by Funda{\c c}{\~a}o para a Ci{\^e}ncia e a Tecnologia (FCT) through national funds (UID/FIS/04434/2013) and Investigador FCT contract IF/01654/2014/CP1215/CT0003., and by FEDER through COMPETE2020 (POCI-01-0145-FEDER-007672).
RB acknowledges support from the ERC advanced grant 339659-MUSICOS.
SC gratefully acknowledges support from Swiss National Science Foundation grant PP00P2\_163824.
RAM acknowledges support by the Swiss National Science Foundation.
This research has made use of \textsc{pyqz} \citep{Dopita:2013}, a Python module to derive the ionisation parameter and oxygen abundance of \ion{H}{ii} regions from their strong emission line ratios hosted at \url{http://http://fpavogt.github.io/pyqz}. \textsc{pyqz} relies on \textsc{statsmodel} \citep{Seabold:2010} and \textsc{matplotlib} \citep{Hunter:2007}.
\end{acknowledgements}

\bibliographystyle{aa}
\bibliography{interaction_gr30}

\end{document}